\documentclass[11pt,envcountsame]{llncs}
\usepackage[left=1.4in,right=1.4in,top=1.4in,bottom=1.4in]{geometry}
\usepackage{graphicx}
\usepackage{hyperref}
\usepackage{amsmath}
\usepackage{amsfonts}
\usepackage{tweaklist}
\renewcommand{\enumhook}{\setlength{\topsep}{0pt}%
  \setlength{\itemsep}{0pt}}

\spnewtheorem{alg}[theorem]{Algorithm}{\bfseries}{\itshape}

\newcommand{\cc}[1]{\mathcal{C}_{#1}}
\newcommand{\RG}[1]{\mathcal{G}_{#1}}

\newcommand{\sH}{\mathcal H}
\newcommand{\sR}{\mathcal R}
\newcommand{\NN}{\mathbb{N}}

\newcommand{\A}{\texttt{A}}
\newcommand{\C}{\texttt{C}}

\newcommand{\pc}{\varphi}

\begin{document}
\makeatletter
\providecommand*{\toclevel@author}{0}
\providecommand*{\toclevel@title}{0}
\makeatother

\mainmatter
\title{Viral population estimation using pyrosequencing}
\author{Nicholas Eriksson\inst{1}\fnmsep%
\thanks{Corresponding author: \email{eriksson@galton.uchicago.edu}} 
\and 
Lior Pachter\inst{2}\and 
Yumi Mitsuya\inst{3} \and
Soo-Yon Rhee\inst{3} \and 
Chunlin Wang\inst{3} \and
Baback Gharizadeh\inst{4} \and
Mostafa Ronaghi\inst{4} \and
Robert W. Shafer\inst{3} \and 
Niko Beerenwinkel\inst{5}}
\institute{
Department of Statistics, University of Chicago \and
Departments of Mathematics and Computer Science, UC Berkeley \and
Division of Infectious Diseases, Department of Medicine, Stanford University \and
Genome Technology Center, Stanford University \and
Department of Biosystems Science and Engineering, ETH Z\"urich
}

\toctitle{Viral population estimation}
\tocauthor{Eriksson et al.}

\maketitle

\begin{abstract}
The diversity of virus populations within single infected hosts presents a
major difficulty for the natural immune response as well as for vaccine design 
and antiviral drug therapy.    Recently developed pyrophosphate based 
sequencing technologies (pyrosequencing) can be used for quantifying this
diversity by ultra-deep sequencing of virus samples.  We present computational
methods for the analysis of such sequence data and apply these 
techniques to pyrosequencing data obtained from HIV populations within patients
harboring drug resistant virus strains.  Our main result is the estimation of 
the population structure of the sample from the pyrosequencing reads.  This
inference is based on a statistical approach to error correction, followed by a 
combinatorial algorithm for constructing a minimal set of haplotypes that 
explain the data.  Using this set of explaining haplotypes, we apply a 
statistical model to infer the frequencies of the haplotypes in the population 
via an EM algorithm.  We demonstrate that pyrosequencing reads allow for effective
population reconstruction by extensive simulations and by comparison 
to 165 sequences obtained directly from clonal sequencing of four independent, 
diverse HIV populations. 
Thus, pyrosequencing can be used for cost-effective estimation of the structure 
of virus populations, promising new insights into viral evolutionary dynamics 
and disease control strategies. 
\end{abstract}

\section*{Synopsis}
The genetic diversity of viral populations is important for biomedical problems
such as disease progression, vaccine design, and drug resistance, yet it is not
generally well-understood.  In this paper, we use pyrosequencing, a novel DNA
sequencing technique, to reconstruct viral populations.  Pyrosequencing
produces a large number of short, error-prone DNA reads.  We develop
mathematical and statistical tools to correct errors and assemble the reads
into the different viral strains present in the population.  We apply these
methods to HIV-1 populations from drug resistant patients and show that
pyrosequencing produces results quite close to accepted techniques at a low
cost and potentially higher resolution.

\newpage

\section{Introduction}

Pyrosequencing is a novel experimental technique for determining the sequence
of DNA bases in a genome \cite{Fakhrai-Rad2002,Margulies2005}.  The method is
faster, less laborious, and cheaper than existing technologies, but pyrosequencing reads are
also significantly shorter and more error-prone (about 100--250 base pairs and
5-10 errors/kb) than those obtained from Sanger sequencing (about 1000 base
pairs and $0.01$ errors/kb) \cite{Malet2002,Huse2007,Wang2007}.

In this paper we address computational issues that arise in applying this
technology to the sequencing of an RNA virus sample.  Within-host RNA virus
populations consist of different haplotypes (or strains) that are
evolutionarily related. The population can exhibit a high degree of genetic
diversity and is often referred to as a quasispecies, a concept that originally
described a mutation-selection balance \cite{Eigen1989,Domingo1997}.  Viral
genetic diversity is a key factor in disease progression
\cite{Nowak1991,Shankarappa1999}, 
vaccine design \cite{Gaschen2002,Douek2006}, 
and antiretroviral drug therapy
\cite{Beerenwinkel2005h,Rhee2007}.  
Ultra-deep sequencing of mixed virus samples is a promising
approach to quantifying this diversity and to resolving the viral
population structure \cite{O'Meara2001,Simons2005,Tsibris2006}. 

Pyrosequencing of a virus population produces many reads, each of which
originates from exactly one---but unknown---haplotype in the population.  
Thus, the central problem is to reconstruct from the read data 
the set of possible haplotypes that is consistent with the observed
reads and to infer the structure of the population, i.e., the relative
frequency of each haplotype.

\begin{figure}[tb]
\centering
\includegraphics[width=0.5\textwidth,angle=270]{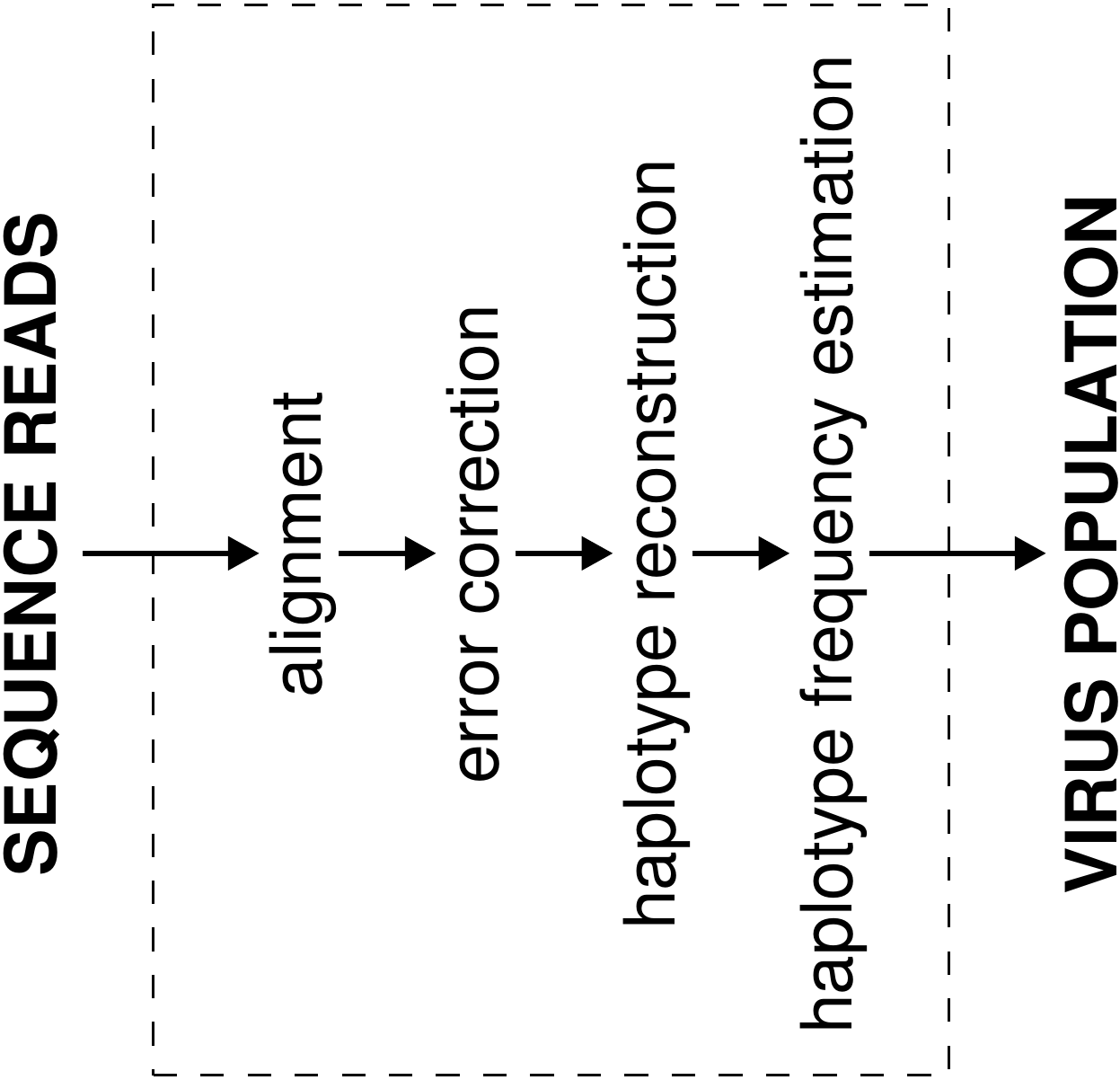}
\caption{Overview of viral population estimation using pyrosequencing. 
Sequence reads are first aligned to a reference strain, then
corrected for errors, and assembled into haplotype candidates.
Finally, the relative frequencies of the reconstructed haplotypes are
estimated in a ML fashion. These estimates constitute the inferred
virus population.
}
\label{fig:overview}
\end{figure}

Here we present a computational four-step procedure for making
inference about the virus population based on a set of pyrosequencing reads
(Fig.~\ref{fig:overview}). First, the reads are aligned to a reference
genome. Second, sequencing errors
are corrected locally in windows along the multiple alignment 
using clustering techniques. Next, we assemble haplotypes that are 
consistent with the observed reads. We formulate this problem as a search 
for a set of covering paths in a directed acyclic graph and show how 
the search problem can be solved very efficiently. Finally, we introduce a 
statistical model that mimics the sequencing process and 
we employ the maximum likelihood (ML) principle for estimating the 
frequency of each haplotype in the population. 

The alignment step of the proposed procedure is straightforward for the data
analyzed here and has been discussed elsewhere \cite{Wang2007}.  Due to the presence of
a reference genome, only pairwise alignment is necessary between each read and the reference genome.
We will
therefore focus on the core methods of error correction, haplotype
reconstruction, and haplotype frequency estimation.  Two independent approaches
are pursued for validating the proposed method.  First, we present extensive
simulation results of all the steps in the method.  Second, we validate the
procedure by reconstructing four independent HIV populations from
pyrosequencing reads and comparing these populations to the results of clonal
Sanger sequencing from the same samples. 

These datasets consist of approximately 5000 to 8000 reads of average length
105 bp sequenced from a 1kb region of the \emph{pol} gene from clinical samples
of HIV-1 populations.
Pyrosequencing (with the Roche GS20 platform \cite{RocheGS20}) can produce
up to 200,000 usable reads in a single run.  Part of our contribution
is an analysis of the interaction between the number of reads, the sequencing
error rate and the theoretical resolution of haplotype reconstruction.
The methods developed in this paper scale to these huge datasets under
reasonable assumptions.  However, we concentrate mainly on a sample size (about
10,000 reads) that produces finer resolution than what is typically obtained
using limiting dilution clonal sequencing.  Since many samples can be run
simultaneously and independently, this raises the possibility of 
obtaining data from about 20 populations with one pyrosequencing run.

Estimating the viral population structure from a set of reads is, in general,
an extremely hard computational problem because of the huge number of possible
haplotypes.  The decoupling of error correction, haplotype reconstruction, and
haplotype frequency estimation breaks this problem into three smaller and more
manageable tasks, each of which is also of interest in its own right.  The
presented methods are not restricted to RNA virus populations, but apply
whenever a reference genome is available for aligning the reads, the read
coverage is sufficient, and the genetic distance between haplotypes is large
enough.  Clonal data indicates that the typical variation in the HIV {\em pol}
gene is about 3-5\% in a single patient \cite{Bacheler2001}.  We find that as
populations grow more diverse, they become easier to reconstruct.  Even at 3\%
diversity, we find that much of the population is reconstructible using our
methods.

The \emph{pol} gene has been sequenced extensively and (essentially) only one
specific insertion seems to occur: the ``69 insertion complex,'' which occurs
under NRTI pressure \cite{Johnson2006}.  None of our samples were treated with NRTIs,
and the Sanger clones did not display this (or any) indel.  Therefore we assume
throughout that there are no true indels in the population.  However, the
algorithms developed in this paper generalize in a straightforward manner for
the case of true indels.

The problem of estimating the population structure from sequence reads is
similar to assembly of a highly repetitive genome \cite{Pevzner2001}.
However, rather than reconstructing one
genome, we seek to reconstruct a population of very similar genomes.  As such,
the problem is also related to environmental sequencing projects, which try to
assess the genomes of all species in a community \cite{Tyson2005}.  While the
associated computational biology problems are related to those that appear in
other metagenomics projects \cite{Chen2005}, novel approaches are required to
deal with the short and error-prone pyrosequencing reads and the complex
structure of viral populations.  The problem is also similar to the haplotype
reconstruction problem \cite{HapMap2005}, with the main difference being that
the number of haplotypes is unknown in advance, and to estimating the diversity
of alternative splicing \cite{Jenkins2006}. 

More generally, the problem of estimating diversity in a population
from genome sequence samples, has been studied extensively for microbial
populations.  For example, the spectrum of contig lengths has been used
to estimate diversity from shotgun sequencing data \cite{Breitbart2002}.
Using pyrosequencing reads, microbial diversity has been assessed by counting 
BLAST hits in sequence databases \cite{Sogin2006}.
Our methods differ from previous work in that we show how to analyze highly
directed, ultra-deep sequencing data using a
rigorous mathematical and statistical framework.

\section{Results}  

We have developed a computational and statistical procedure for
inferring the structure of a diverse virus population
from pyrosequencing reads. Our approach comprises four 
consecutive steps (Fig.~\ref{fig:overview}),
starting with the alignment of reads to a reference sequence and
followed by error correction, haplotype reconstruction, and
haplotype frequency estimation.

\subsection{Error correction}   \label{sec:error}

Given the high error rate of pyrosequencing, error correction is a necessary
starting point for inferring the virus population.  The errors in
pyrosequencing reads typically take the form of one-base indels along with
substitutions and ambiguous bases and occur most often in 
homopolymeric regions.
The reads come with quality scores for each base quantifying the
probability that the base is correct.

Error rates with the Roche GS20 system have been estimated as approximately
5--10 errors per kb \cite{Huse2007,Wang2007}.  However, a small number of reads
accounts for most of the errors.  Thus after discarding approximately 10\% of
the reads (those with ambiguous bases or atypical length), the error can be
reduced to 1--3 errors per kb \cite{Huse2007}.  These remaining errors are
about half insertions and a quarter each deletions and substitutions.

Due to our assumption that there are no haplotypes with insertions in the
population, the insertions in the reads can all be simply corrected through
alignment with the reference genome.  We do not deal with the problem of
alignment here; it is straightforward because our assumption of the
existence of a reference genome implies that only
pairwise alignment is necessary.
In the remainder of the paper, we assume that a correct alignment is given,
leaving about 1 error per kb to correct (or 3 errors per kb if the low-quality
reads are not aggressively pruned).

%

\begin{figure}[tb]
\centering
\includegraphics[width=0.3\textwidth,angle=270]{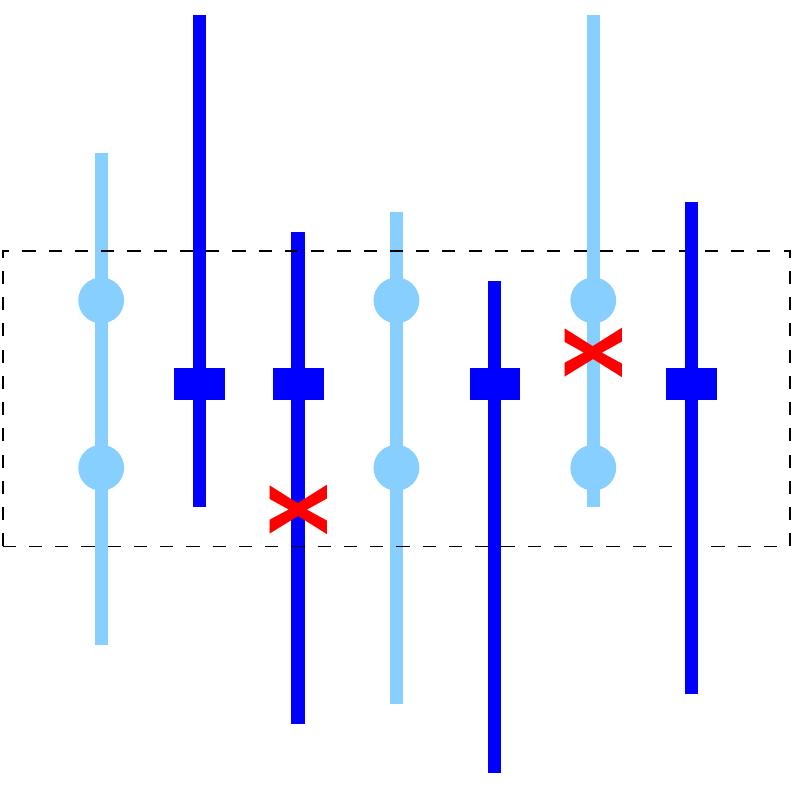}
\caption{Error correction. Fixed-width windows (shown as the dashed box) over
the aligned reads are considered. Two different types of reads are depicted
(light versus dark lines) indicating their origin from two different
haplotypes.  Genetic differences (indicated by circles and squares) provide the
basis for clustering reads into groups representing the respective haplotypes.
After clustering, errors (marked as crosses) can be corrected.}
\label{fig:clustering}
\end{figure}

Our approach for error correction resembles the method of
\cite{Kececioglu2001} for
distinguishing repeats in whole genome shotgun assemblies combined
with \cite{Tammi2002}. We consider all the reads in a window over the multiple
alignment and cluster these reads using a statistical testing procedure to
detect if a group of reads should be further split (Fig.~\ref{fig:clustering}).
The reads in each cluster are then corrected to the consensus sequence of the
cluster.

The statistical testing procedure consists of two steps.  First, every column
in the window is tested for over-representation of a mutation using a binomial
test.  Second, every pair of columns is tested to see if a pair of mutations
happen together more often than would be expected by chance.  See
Section~\ref{sec:ecmethod} for details on the tests.

Any significant over-representation of a mutation or a pair in a window is
regarded as evidence for the reads originating from more than one haplotype.
The testing procedure produces an estimate for the number of haplotypes in the
window as follows.  First, all single mutations are tested for significance;
each significant mutation gives evidence for another haplotype in the window.
Next, all pairs of mutations are tested; any significant pairs is evidence for
another haplotype.  However, this process can over-count the number of
haplotypes in the window in certain cases if two mutations are significant both
by themselves and as a pair.  In this case, we correct for the over-count, see
Section~\ref{sec:ecmethod}.

We then separate the reads into $k$ groups using 
$k$-means clustering.
The algorithm is initialized with both random clusters and
clusters found by a divisive clustering method based on the statistical tests.
We use the Hamming distance
between sequences to calculate cluster membership;
the consensus sequences define the cluster centers. The consensus sequence
is computed from weighted counts based on the quality scores. 
Thus, the inferred cluster centers are the reconstructed haplotypes.
Combining testing and clustering we proceed as follows:
\begin{alg}
        \label{alg:ec}
{\bf (Local error correction)}  

\noindent {\em Input:} 
A window of aligned reads.

\noindent {\em Output:}
The $k$ haplotypes in the window and the error corrected reads.
 
\noindent {\em Procedure:}
\begin{enumerate}

\item Find all candidate mutations and pairs of mutations and test for
        overrepresentation.

\item Count the number of non-redundant mutations and pairs that are significant.  
        This is the number $k$ of  haplotypes in the window.

\item Cluster the reads in the window into $k$ clusters and correct each read to its cluster center.

\item Output corrected reads.   
\end{enumerate} 
\end{alg}

Applying a parsimony principle the algorithm finds the smallest number $k$ of
haplotypes that explain the observed reads in each window. The genomic region to
be analyzed is divided into consecutive windows and Algorithm~\ref{alg:ec} is
run in each of them. We use three collections of successive windows that are
shifted relative to each other such that each base in the region is covered
exactly three times. The final correction of each base is the consensus of the
three runs. 

The error correction procedure can lead to uncorrected errors or miscorrections
via false positives and negatives (leading to over/underestimation of the
number of haplotypes in a window) or misclustering.
See Section~\ref{sec:ecmethod} for implementation details and
a discussion of setting the parameters so as to minimize these mistakes.

False positives arise when an error is seen as a significant variant;
they will be consequences of setting the error rate too low 
or the significance level too high, or if errors are highly correlated.
Misclustering can happen if errors occur frequently enough on 
a single read to make that read appear closer to an incorrect haplotype.
This likelihood is increased as the window size grows and more reads 
overlap the window only partially.

An analysis of the false negative rate gives an idea of the theoretical
resolution of pyrosequencing.  False negatives arise when a true variant tests
as non-significant and thus is erased. If the input data was error-free, this
would be the only source of mistaken corrections and would happen by
eliminating rare variants.  Given an error rate of 2.5 errors per kb, the 
calculation in Section~\ref{sec:ecmethod} shows that variants present in under
1\% of the population would be erased on a dataset of 10,000 reads.  In
Section~\ref{sec:haplo} below, we will show that this number of reads is about
enough to expect to resolve haplotypes present in 1\% of the population.

\subsection{Haplotype reconstruction} \label{sec:haplo}

Our approach to haplotype reconstruction rests on two basic beliefs.  First, the
haplotypes in the populations should not exhibit characteristics that are not
present in the set of reads.  This means that every haplotype in the population
should be realizable as an overlapping series of reads.  Second, the population
should explain as many reads as possible with as few haplotypes as possible.  

We assume a set $\sR$ of aligned and error-corrected reads
obtained from sequencing a population.  
If all haplotypes have the same length $n$, then
each aligned read consists of a start position in the genome and a string
representing the genomic sequence.  
We say that two reads overlap if there are
positions in the genome to which they are both aligned.  They agree on their
overlap if they agree at all of these positions.
We call a haplotype \emph{completely consistent} with the set of reads $\sR$ 
if the haplotype can be constructed from a subset of overlapping reads of
$\sR$ that agree on their overlaps.  Let
$\cc{\sR}$ be the set of all haplotypes that are completely consistent with
$\sR$.  In the following, we provide methods for constructing and sampling
from $\cc{\sR}$ and we present an efficient algorithm for computing a lower 
bound on the number of haplotypes necessary to explain the reads.
Both techniques rely on the concept of a read graph.



\begin{definition}   \label{def:readgraph}
{\bf (Read graph)}
The \emph{read graph} $\RG{\sR}$ associated with a set of reads $\sR$ is the
acyclic directed graph with vertices $\{\sR_{\rm irred}, s, t\}$ consisting of
a source $s$, a sink $t$, and one vertex for every irredundant read 
$r \in \sR$. Here, a read is \emph{redundant} if there is
another read that overlaps it completely such that the two reads agree on
their overlap.  
The edge set of $\RG{\sR}$ is defined by including
an edge from an irredundant read $r_1$ to an
irredundant read $r_2$, if
\begin{enumerate}
\item $r_1$ starts before $r_2$ in the genome,
\item $r_1$ and $r_2$ agree on their (non-empty) overlap, and
\item there would not be a path in $\RG{\sR}$ from $r_1$ to $r_2$ without this edge.
\end{enumerate}
Finally, edges are added from the source $s$ to all reads beginning at
position 1, and from all reads ending at position $n$ to the sink $t$.
\end{definition}

A path in the read graph from the source to the sink corresponds to  
a haplotype that is completely consistent with $\sR$.
Thus, finding
$\cc{\sR}$, the set of completely consistent haplotypes, 
amounts to efficiently enumerating paths in the read graph.  

\begin{figure}[tb]
\begin{center}
        \begin{tabular}{c@{\hspace{1.5cm}}c}
\includegraphics[width=.12\textwidth]{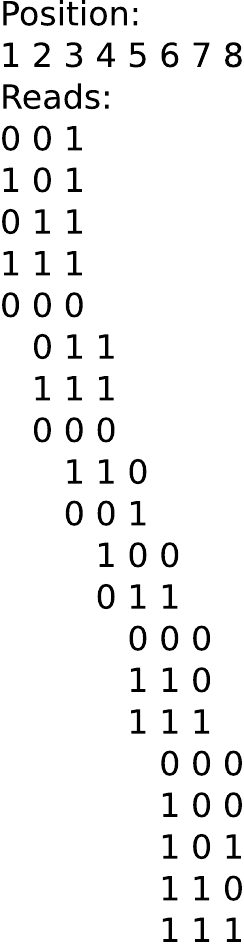} & 
\includegraphics[width=.7\textwidth]{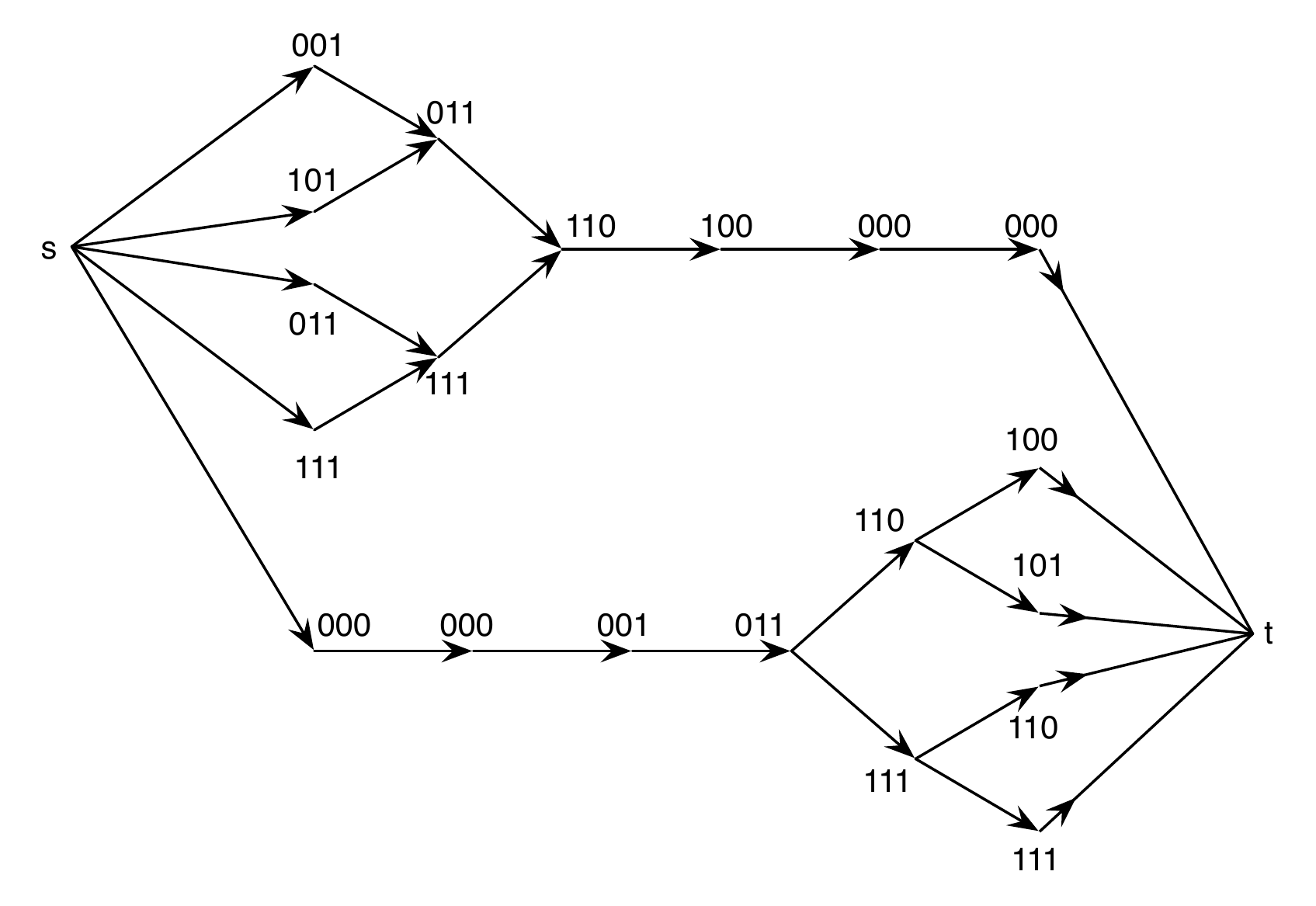}\\
~ & ~\\
(a) & (b)
\end{tabular}
\caption{Read graph. A simplified genome of length $n=8$ over the binary
  alphabet $\{0,1\}$ is considered. Twenty reads of length 3 each are
  aligned to an assumed  reference sequence (a). The induced read graph
  has $20+2$ vertices and 28 edges (b).
}
\label{fig:readgraph}
\end{center}
\end{figure}

For example, in Fig.~\ref{fig:readgraph} a simplified 
genome of length $n=8$ over the binary alphabet $\{0,1\}$
is considered, and an alignment of $20$ reads (each of length 3) is 
shown in Fig.~\ref{fig:readgraph}a. These data give rise
to the read graph depicted in Fig.~\ref{fig:readgraph}b.
For instance, the haplotype {\tt 00110000} is completely consistent with the
reads and corresponds to the top path in the graph.  For more complex
read graphs, see Fig.~\ref{fig:RG}.

We say that a set of haplotypes $\sH$ is an \emph{explaining}
set for $\sR$ if every read $r \in \sR$ can be obtained as a substring of some haplotype
in $\sH$. 
We seek a small set of explaining haplotypes 
and focus on the set $\cc{\sR}$, which consists exactly of those
haplotypes that emerge from the data. 
The following proposition provides a criterion for $\cc{\sR}$ to
be an explaining set in terms of the read graph.
\begin{proposition}\label{prop:cc}
The set of haplotypes completely consistent with a set of reads
is an explaining set for these reads if, and only if, every vertex
of the read graph lies on a directed path from the source to the sink.
\end{proposition}

The Lander--Waterman model of sequencing is based 
on the assumptions that reads are random 
(uniformly distributed on a genome) and independent \cite{Lander1988}.
In this model, the probability that all bases of a genome
of length $n$ are sequenced follows the Poisson distribution
$p = (1 - e^{-c})^n$, where $c$ is the coverage (the total number
of bases sequenced per position).
For a sequencing experiment from a mixed population with
different abundances of haplotypes (or subspecies),
a similar approach can be applied \cite{Chen2005}.
For the probability of complete coverage of all haplotypes
occurring with a frequency of at least $\rho$, we have
$p \geq (1 - e ^{-c \cdot \rho})^{n}$.
Since $c = NL/n$, where $N$ is the number of reads and $L$ is
the read length, sequencing
\begin{equation}  \label{eq:LW}
   N \geq - \frac{n \, \ln(1 - p^{1/n})}{\rho \, L}
\end{equation}
reads will ensure that the completely consistent haplotypes 
assembled from the reads are an explaining set for these haplotypes.
For example, in order to cover all haplotypes of 5\% or higher frequency
of length 1000 bases with reads of length 100 with 99\% probability, 
at least 2302 reads need to be sequenced; to reach haplotypes at 1\% frequency
with 99\% probability, 11,508 reads are needed.
Notice that the number of reads needed scales linearly with the the inverse
of the smallest frequency desired.
We note that the actual number of required reads can be much smaller in
genomic regions of low diversity.

If Proposition~\ref{prop:cc} is violated, we can remove the violating set of
reads to obtain a new set satisfying Proposition~\ref{prop:cc}.  This amounts
to discarding reads that either contain mistakes in the error correction or
come from haplotypes that are at a too low frequency in the
population to be fully sequenced.
Thus the resolution is inherently a function of the number of reads.


We are now left with finding a minimal explaining set of completely 
consistent haplotypes. Restricting to this subset of haplotypes
reduces the computational demand of the problem significantly.
Proposition~\ref{prop:cc} implies that an explaining set of completely
consistent haplotypes is precisely a set of paths in the read graph 
from the source to the sink, such that all vertices of the read graph
are covered by at least one path.  We call such a set of paths a
\emph{cover} of the read graph.  The following result shows that a minimal
cover can be computed efficiently (see Methods for a proof).

\begin{theorem}
\label{thm:dilworth}
{\bf (Minimal cover of the read graph)}
\vspace{-1ex}
\begin{itemize}
\item[\rm (1)] Every minimal cover of the read graph has the same cardinality, namely
the size of the largest set $Q$ of vertices such that there
are no paths between elements of $Q$. 
\item[\rm (2)] A minimal cover of the read graph can be computed by solving a maximum
matching problem in an associated bipartite graph.   This matching problem
can be solved in time at worst cubic in the number of irredundant reads.
\end{itemize}
\end{theorem}

The minimal path cover obtained from the maximum matching algorithm is
in general not unique.  First, it provides a minimal chain decomposition of the
graph.  A \emph{chain} in a directed acyclic graph is a set of vertices that all
lie on at least one common path from the source to the sink.   A chain can
generally be extended to a number of different paths.
Second, the minimal chain decomposition itself is in general not unique.  
However, the cardinality of the minimal cover is well-defined.
It is an important invariant of the set of
reads, indicating the smallest number of haplotypes that can explain the data.
Notice that the size of the minimal read graph cover 
can be greater than the maximum number of haplotypes in a given window
of the error correction step. The cardinality of the minimal cover
is a global invariant of the set of reads.

\begin{alg}
\label{alg:min}
{\bf (Construction of a minimal set of explaining haplotypes)}  

\noindent {\em Input:} 
A set $\sR$ of aligned, error corrected reads satisfying the conditions of Prop.~\ref{prop:cc}.

\noindent {\em Output:}
A minimal set of explaining haplotypes for $\sR$.
 
\noindent {\em Procedure:}
\begin{enumerate}
        \item Construct the read graph $\RG{\sR}$ associated with $\sR$.
        \item Compute a minimal chain decomposition of the read graph.
        \item Extend the chains in the graph to paths from the source to the
              sink in $\RG{\sR}$.
        \item Output the set of haplotypes corresponding to the paths found in step~3.
\end{enumerate}
\end{alg}

The algorithm can easily be modified to produce a non-minimal set by
constructing multiple chain decompositions and by choosing multiple ways to
extend a chain to a path.  We note that the set of all paths in the graph is
generally much too large to be useful. For example, the HIV datasets
give rise to up to $10^9$ paths and in simulations we often found over $10^{12}$
different paths in the graph.  Generating paths from minimal explaining sets is
a reasonable way of sampling paths, as we will see in Section~\ref{sec:sim}
(also see Fig.~\ref{fig:RG} and \ref{fig:multCD}).

Finally, if the condition of Proposition~\ref{prop:cc} is not satisfied,
i.e., if the coverage is too low and the set of completely consistent haplotypes
does not contain an explaining set, then that condition can be relaxed.
This corresponds to modifying the read graph by adding edges between all
non-overlapping reads. Algorithm~\ref{alg:min} will then again find
a minimal set of explaining haplotypes.

\subsection{Haplotype frequency estimation} \label{sec:em}

A virus population is a probability distribution on a set of haplotypes.  We
want to estimate this distribution from a set of observed reads.  Let $\sH$ be
a set of candidate haplotypes.  In principle, we would like $\sH$ to be the
set of all possible haplotypes, but in practice
we must restrict $\sH$ to a smaller set of explaining
haplotypes as derived from Algorithm~\ref{alg:min} in order to make the
estimation process feasible.  Let $\sR$ be the set of all possible reads
that are consistent with the candidate haplotypes in $\sH$.  The read data is
given as a vector $u \in \NN^\sR$, where $u_r$ is the number of times that read
$r$ has been observed. 

\begin{figure}[tb]
\centering
\includegraphics[width=0.2\textwidth,angle=270]{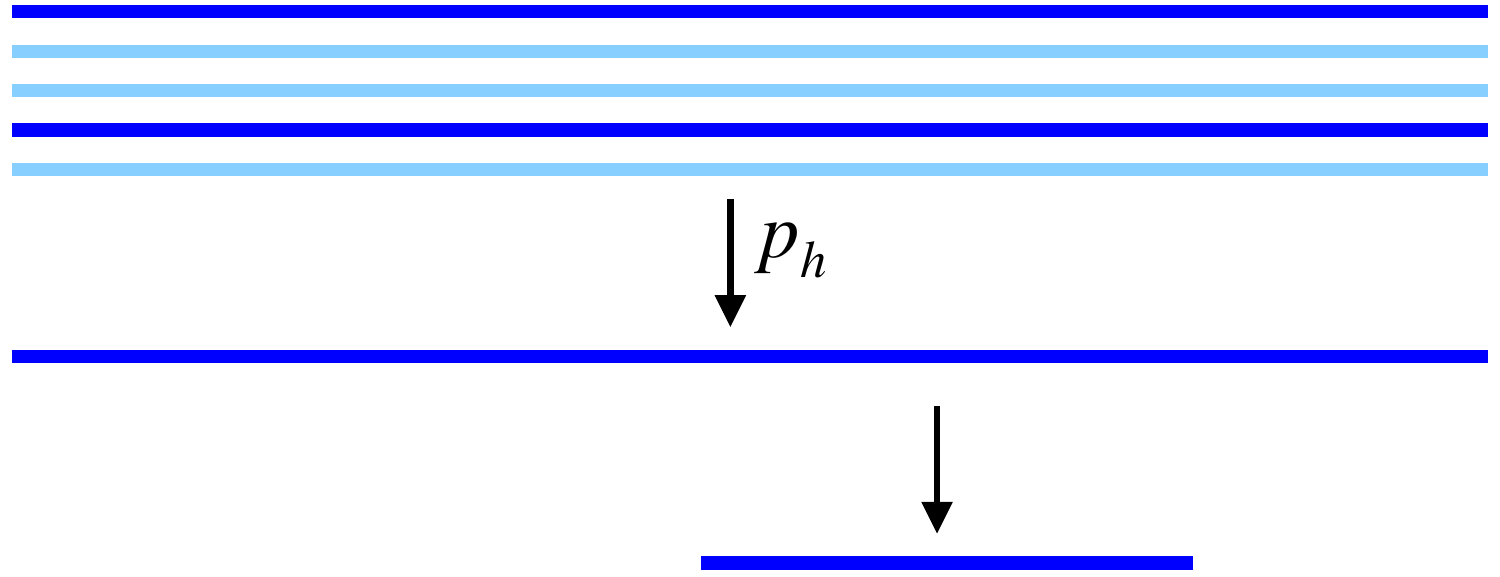}
\caption{Schematic representation of the sampling process.
The virus population is represented by five genomes
(top) of two different haplotypes (light versus dark lines).
The probability distribution is $p = (3/5, \, 2/5)$. The
generative probabilistic model assumes that haplotypes are
drawn from the population according to $p$, 
and reads are sampled uniformly from the haplotypes (bottom).
}
\label{fig:sampling}
\end{figure}

Our inference is based on a statistical model for the generation
of sequence reads from a virus population.
Similar models have been used for haplotype frequency estimation
\cite{Excoffier1995,Stephens2001,Halperin2006}.
We assume that reads are sampled as follows (Fig.~\ref{fig:sampling}). 
First, a haplotype $h$ is drawn at
random from the unknown probability distribution $p = (p_h)_{h \in \sH}$. 
Second, a read $r$ is
drawn with uniform probability from the set of all reads
with which the haplotype is consistent. 
Estimating the structure of the population is the
problem of estimating $p$ from $u$ under this generative model.

Let $H$ be the hidden random
variable with values in $\sH$ that 
describes the haplotype and $R$ the observed random variable
over $\sR$ for the read. Then the
probability of observing read $r$ under this model is
\[
   \Pr(R = r) = \sum_{h \in \sH} p_h \Pr(R = r \mid H = h),
\]
where the conditional probability is defined as
$\Pr(R = r \mid H = h) = 1/K$ if $h$ is consistent with $r$,
and 0 otherwise.
Here $K$ is the number of reads $r \in \sR$ that $h$
is consistent with.  
Since we assume that all haplotypes have the same length,
$K$ is independent of both $r$ and $h$.

We estimate $p$ by maximizing the log-likelihood function
\[
   \ell(p_1, \dots, p_{|\sH|}) = \sum_{r \in \sR} u_r \log \Pr(R = r).
\] 
This is achieved by employing an EM algorithm (see
Section~\ref{sec:emmethod} for details). 
Each iteration of the EM algorithm runs in time $O(|\sR||\sH|)$.  
For example, for 5000 reads and 200 candidate haplotypes, 
the EM algorithm typically converges within minutes
on a standard PC.

Software implementing the algorithms for error correction, haplotype
reconstruction, and frequency estimation is available upon request from the
authors.


\subsection{Simulation results} \label{sec:sim}


We have simulated HIV populations of different diversities
and then generated reads from these populations by simulating 
the pyrosequencing procedure with various error rates 
and coverage depths. 
The first 1kb of the HIV {\em pol}
gene was the starting point for all simulations.  
We separately analyze the performance 
first of error correction, 
then of haplotype reconstruction,
then of haplotype frequency estimation, 
and finally of the combination of these three steps.

The simulations show that Algorithm~\ref{alg:ec} 
reduces the error rate by a factor of 30. This performance
is largely independent of the number of haplotypes in 
the population (Fig.~\ref{fig:ec}).
ReadSim \cite{Schmid2006} 
was used to simulate the error process of
pyrosequencing. 
The error rate after alignment is
about 1--3 errors per kb, so 
we are left with about 0.1 errors per kb after error correction.
As the population grows and
becomes more diverse, the alignment becomes more difficult
resulting in a smaller error reduction (Fig.~\ref{fig:ec}).

In order to assess the ability of Algorithm~\ref{alg:min} to
reconstruct 10 haplotypes from 10,000 error-free reads
(yielding about 1500 irredundant reads), we generate increasing
numbers of candidate haplotypes. 
This is achieved by repeatedly
finding a minimal set of explaining haplotypes (see Section~\ref{sec:haplo})
until either we reach the desired number of haplotypes or we are unable
to find more haplotypes that are part of a minimal explaining set.
\begin{figure}[tb]
        \centering
        \begin{tabular}{c@{\hspace{1cm}}c}
                \includegraphics[width=.45\textwidth]{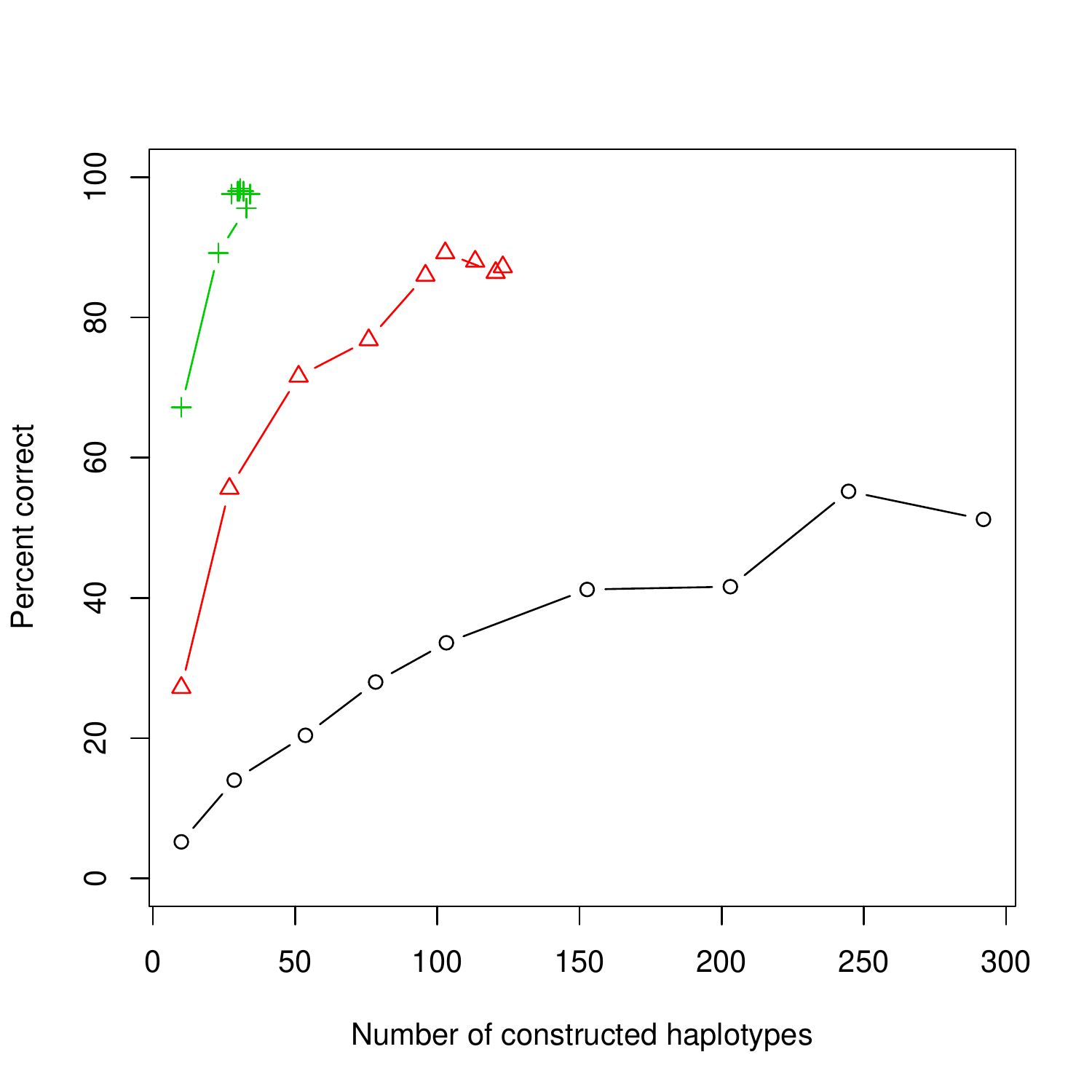} & 
                \includegraphics[width=.45\textwidth]{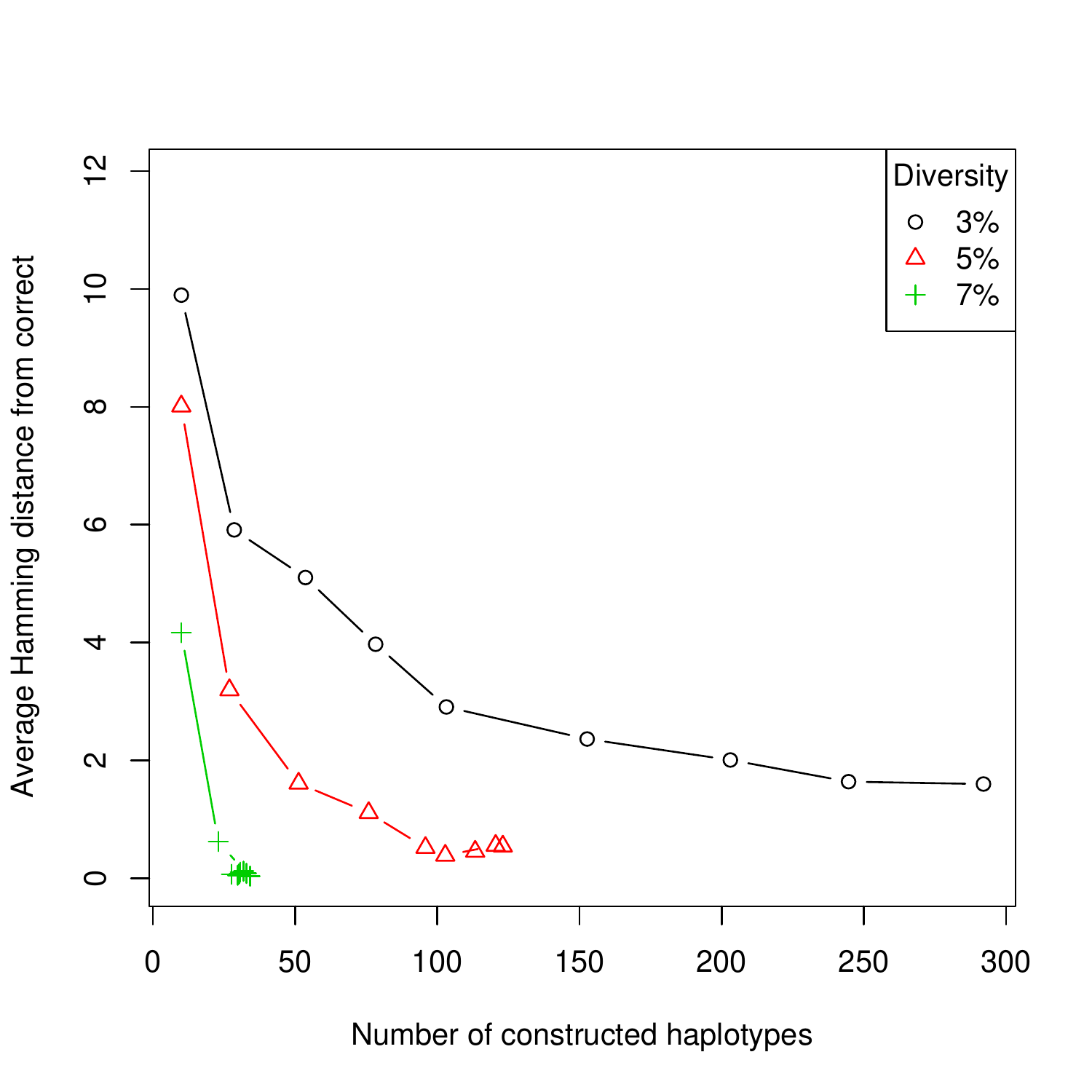}\\
                (a) & (b) \\
        \end{tabular}
		\caption{Haplotype reconstruction.  Up to 300 candidate haplotypes were
		generated using Algorithm~\ref{alg:min} from 10,000 error free reads
		drawn from populations of size 10 at varying diversity levels.
		Displayed are two measures of the efficiency of haplotype
		reconstruction: the percent of the original haplotypes with exact
		matches among the reconstructed haplotypes (a), and the average Hamming
		distance (in amino acids) between an original haplotype and its closest
		match among the reconstructed haplotypes (b).}
        \label{fig:recons} 
\end{figure}
Fig.~\ref{fig:recons} visualizes the enrichment of
recovered true haplotypes with increasing number of
candidate haplotypes for different levels of population diversity.
While in low-diversity populations exact haplotype reconstruction
can be very challenging (Fig.~\ref{fig:recons}a), the algorithm
will always find haplotypes that are close to the true ones. 
For example, at 5\% diversity 10 out of 50 candidate
haplotypes will match the original 10 haplotypes
at an average Hamming distance of just 1.6 amino acid differences 
(Fig.~\ref{fig:recons}b).  With
larger populations, the performance is similar although
more candidate haplotypes need to be generated (Fig.~\ref{fig:20haplo}).
Given that the total number of paths in the read graphs 
considered here is over $3 \cdot 10^{10}$, the strategy of
repeatedly finding minimal sets of explaining haplotypes
is very efficient for haplotype reconstruction.

\begin{figure}
	\centering
	\includegraphics[width=.7\textwidth, height=.252\textheight]{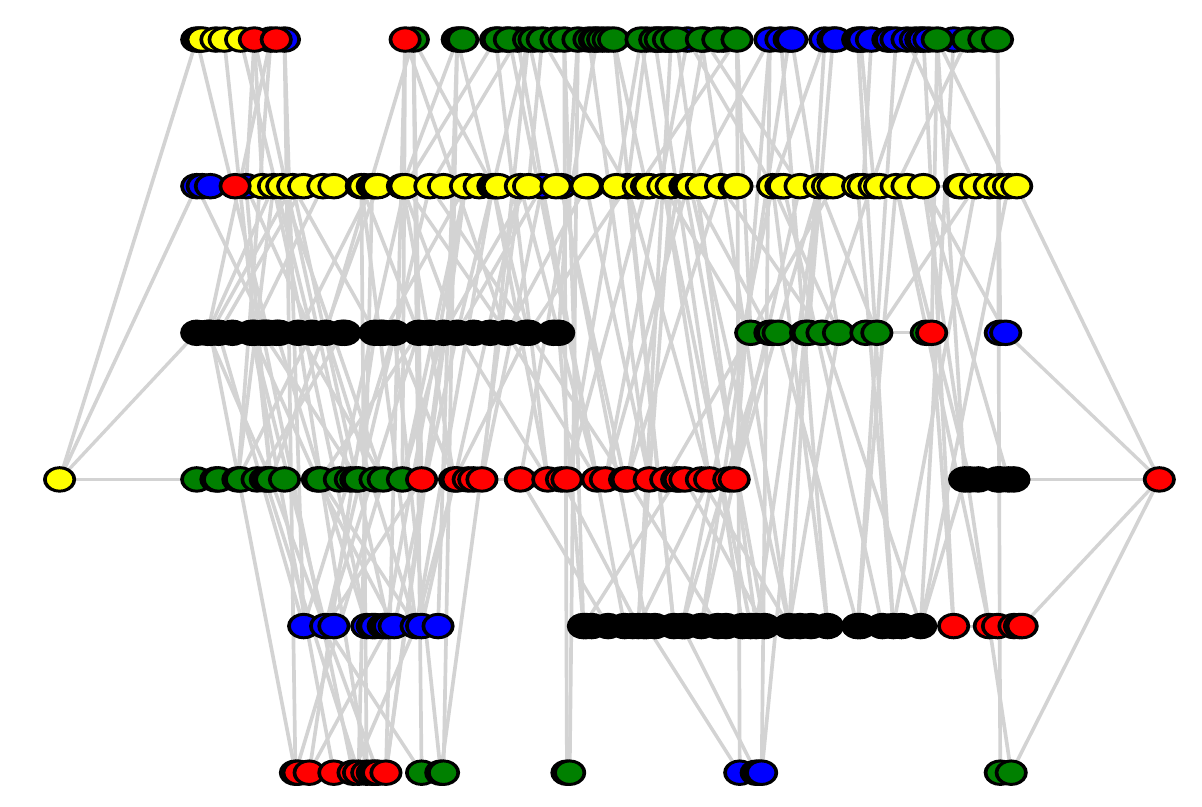} \\(a)
	\bigskip

	\includegraphics[width=.7\textwidth, height=.252\textheight]{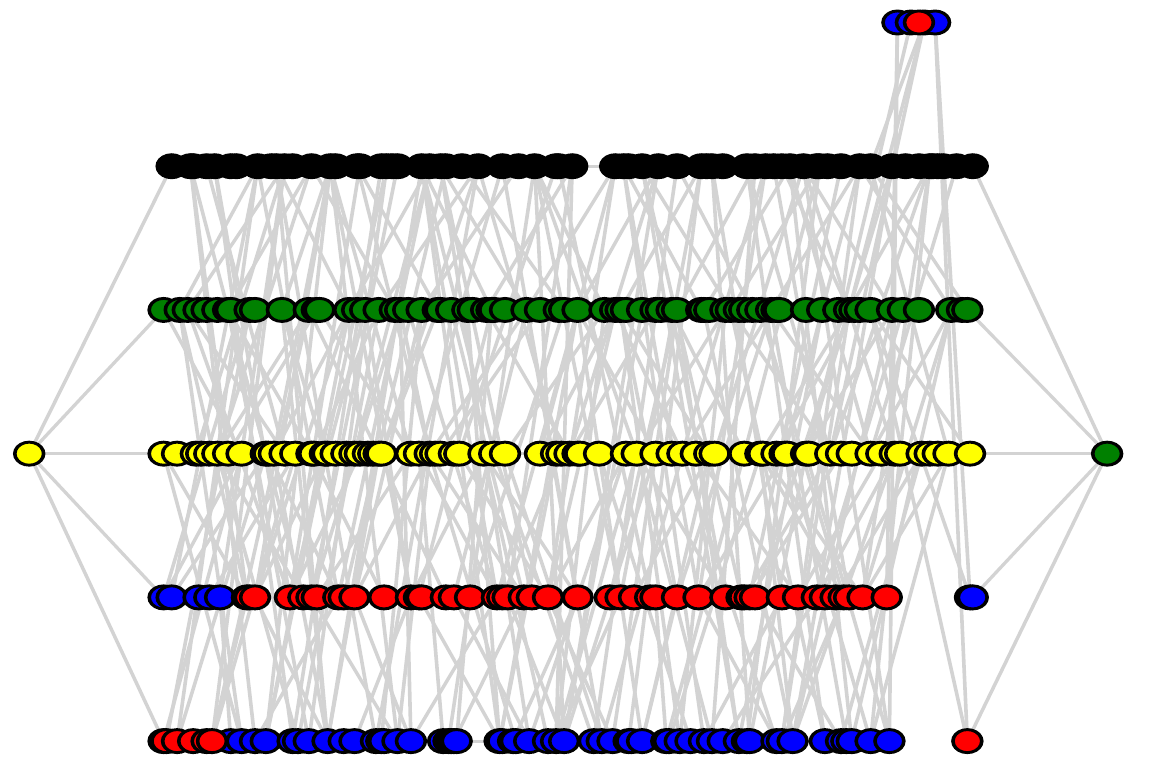} \\(b)
	\bigskip

	\includegraphics[width=.7\textwidth, height=.252\textheight]{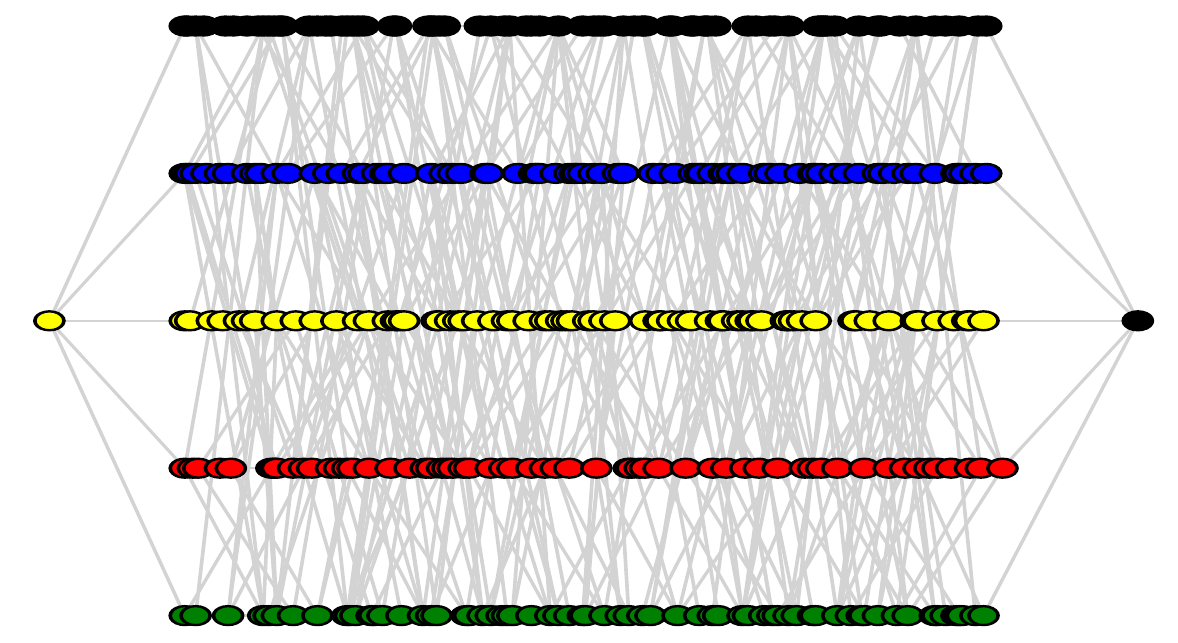}\\ (c)
	\caption{Read graphs for 1000 reads from populations of 5 haplotypes
	at 3\% (a), 5\% (b), and 7\% diversity (c).  
	The bottom five lines in the graph correspond to reads
	which match the five haplotypes uniquely; the top line in subfigures (a)
	and (b) contains reads which match several haplotypes.  In each subfigure,
	the reads are colored according to a single chain decomposition.}
	\label{fig:RG}
\end{figure} 

Fig.~\ref{fig:RG} shows how the haplotype reconstruction problem gets harder at
lower diversity. In each graph, the bottom five lines correspond to reads
matching one of the original five haplotypes uniquely. The sixth line on top
(if present) corresponds to reads which could come from several haplotypes.  At
3\% diversity (Subfigure (a)), only one of the haplotypes is reconstructed
well.  
At 5\% diversity (Subfigure (b)), the decomposition is almost correct except
for a few small ``crossovers''.  
At 7\% diversity (Subfigure (c)), the chain decomposition exactly reconstructs
the five haplotypes.  
By using multiple decompositions we can reconstruct many of the haplotypes
correctly (Fig.~\ref{fig:multCD}) even in low diversities.

\begin{figure}[tb]
        \centering
        \includegraphics[width=.6\textwidth]{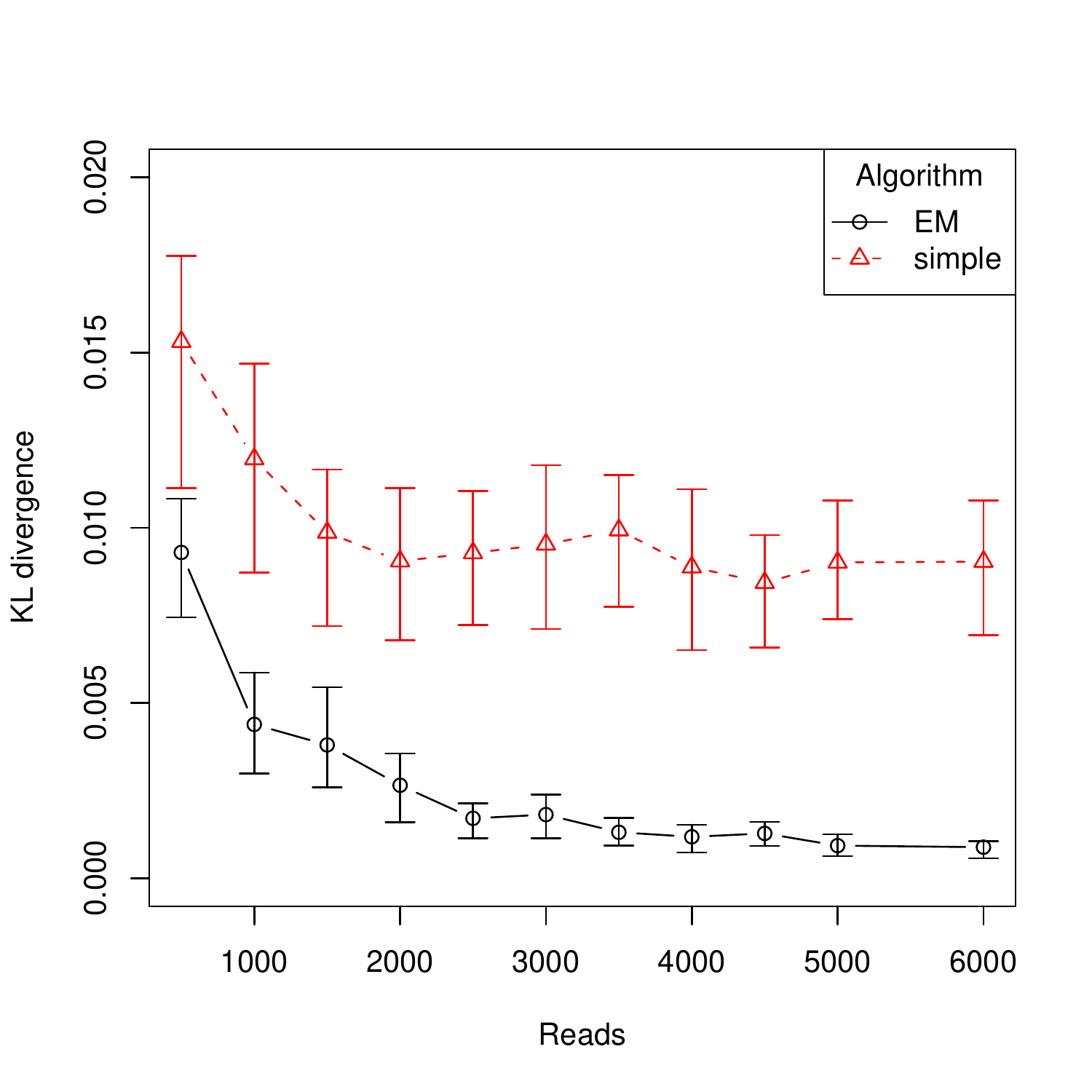}
        \caption{Haplotype frequency estimation.
		Haplotype frequencies were inferred 
		using both the EM algorithm in Section~\ref{sec:em} (circles) 
        and a simple heuristic algorithm (triangles); the resulting
		distance from the correct frequencies 
		is measured using the KL divergence.
		Error bars give the interquartile range over 50 trials.
		The populations consisted of 10 haplotypes at equal frequency and 5\% diversity.
		The input to the algorithms was 
		a set of reads simulated from the population and the original 10 haplotypes.  }
        \label{fig:EM}
\end{figure}

The performance of the EM algorithm for haplotype
frequency estimation described in Section~\ref{sec:em}
is measured as the Kullback--Leibler (KL) divergence
between the original population $p$ and its estimate $\hat{p}$.
We consider populations with 10 different haplotypes, 
each with frequency $1/10$, at 5\% diversity.
Haplotype frequencies are estimated from between 500
and 6000 error-free reads (Fig.~\ref{fig:EM}).  
The performance of the EM algorithm is compared to that of
a simple heuristic method, which assigns frequencies to the 
haplotypes in proportion to the number of reads they explain
(see Methods, Section~\ref{sec:methodsim}).  
For both methods, the KL divergence $D_{\rm KL}(p \parallel \hat{p})$
decreases roughly exponentially with the number of reads.
However, the EM algorithm significantly outperforms the 
heuristic for all sizes of the read set and this improvement 
in prediction accuracy increases with the number of reads.

In order to test the combined performance of the haplotype reconstruction
and frequency estimation, 
our basic measure of performance is 
the proportion of the original population that
is  reconstructed within 10 amino acid differences.  This measure, which we
call $\pc_{10}$, is defined as follows (see also Section~\ref{sec:methodsim}).
For each inferred haplotype, we determine the closest original haplotype
and sum up the frequencies of all inferred haplotypes that differ 
from their assigned original haplotypes by at most ten sites.  
This performance measure indicates how much of the population
has been reconstructed reasonably well.  It is less sensitive to
how well haplotypes and haplotype distributions match 
(see Fig.~\ref{fig:recons} and \ref{fig:EM} for those
performance measures).

\begin{figure}[tb]
        \centering
        \includegraphics[width=.75\textwidth]{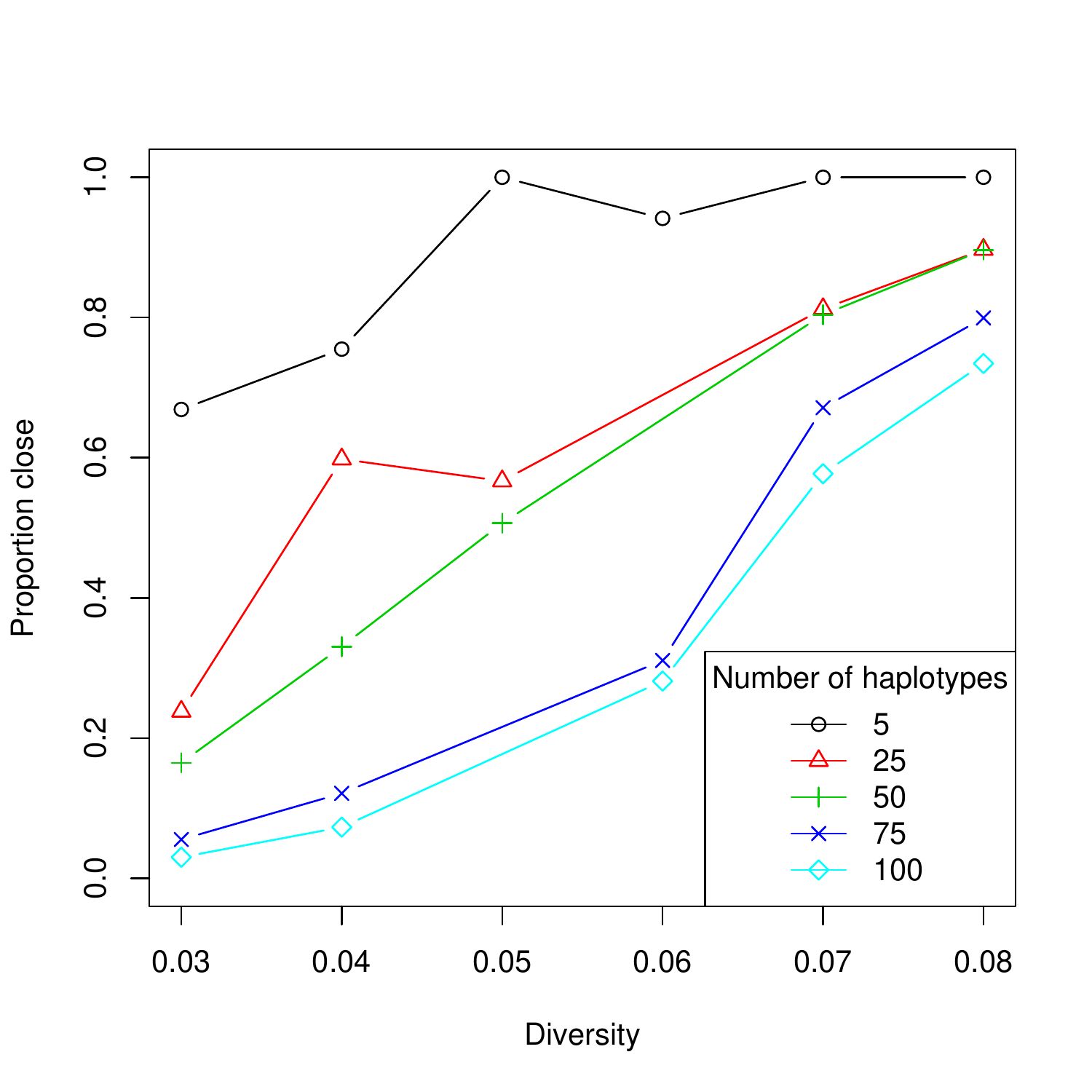}
        \caption{Combined population reconstruction procedure. The
		proportion of the population reconstructed within 10 amino acid
		differences ($\pc_{10}$, ``Proportion close'') is shown.  Here 10,000
		error-free reads were sampled from populations with diversity between 3
		and 8\% and with between 5 and 100 haplotypes of equal frequency.
		Haplotypes were reconstructed and then frequencies were estimated.}
        \label{fig:2dim}
\end{figure}

For the first simulation of combined performance, we consider error-free reads
from populations consisting of between 5 and 100 haplotypes, each with equal
frequency, at diversities between 3 and 8\%.  
We simulated 10,000 error-free reads of average length
100 from these populations and ran haplotype reconstruction and frequency
estimation.  Fig.~\ref{fig:2dim} shows that
performance increases as diversity increases and drops slightly as the number
of haplotypes increases.  As we saw above, we would expect to be able to
reconstruct populations with size approximately 100 using 10,000 reads under
the Lander--Waterman model.

\begin{figure}[tb]
        \centering
        \includegraphics[width=\textwidth]{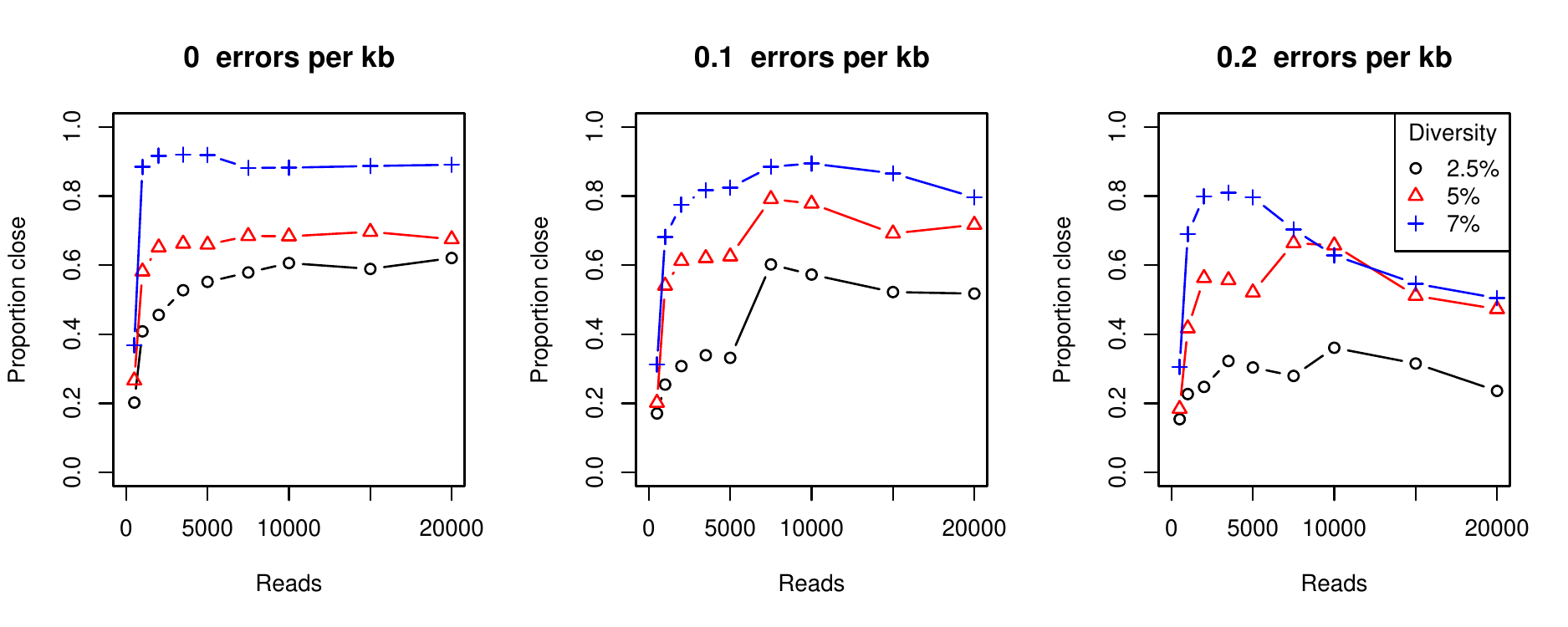}
        \caption{Population reconstruction with errors.
		Proportion of population reconstructed within 10 amino acid differences
		($\pc_{10}$, ``Proportion close'') using haplotype reconstruction and
		frequency estimation.  The original populations had 10 haplotypes of
		equal frequency at varying levels of diversity.  Error was randomly
		introduced in the simulated reads to mimic various levels
		of error correction.}
        \label{fig:close}
\end{figure}

For the second combined test, we tested all three steps: 
error correction, haplotype reconstruction, and frequency estimation.
In order to model the miscorrection of errors, we ran ReadSim \cite{Schmid2006}
to simulate the actual error process of pyrosequencing and then ran error
correction.  New, error-free reads were simulated and  errors were added
through sampling from the distribution of the uncorrected errors
in order to reach error rates of exactly $0.1$ and $0.2$ errors per kb.
Fig.~\ref{fig:close} summarizes the results of this analysis for
10 haplotypes at varying diversities.
The combined procedure performs very well on error-free reads that
are diverse enough.  As errors are introduced, performance
decreases; however the method still recovers much of the original population.
For example, at 0.1 errors per kb, which is the error rate expected with
current pyrosequencing technology and our error correction method (see above), 
as few as 3500 reads are required for approximately recovering 55\% 
of a population of 5\% diversity. 

Fig.~\ref{fig:close} also indicates, for the datasets with error rate $0.2$,
a small performance loss as the number of reads increases.  
This phenomenon appears to be related to the fact that more reads give rise to
more paths in the graph, thereby increasing the chances that completely 
consistent haplotypes that contain errors are assigned positive
probabilities.  In fact, the size of a minimal path cover increases 
approximately linearly with the number of reads and this increase 
does not appear to depend much on population diversity
(Fig.~\ref{fig:antichains}).

\subsection{Analysis of HIV samples}
\label{sec:HIV}

Our second evaluation of population reconstruction is based on ultra-deep
sequencing of drug-resistant HIV populations from four  infected patients
\cite{Wang2007}.
The four virus populations were analyzed independently using pyrosequencing and
clonal Sanger sequencing. 
Table~\ref{tab:results} shows the
resulting statistics on the datasets.  
The pyrosequencing based approach
mirrors very closely the clonal sequencing.  
To compare the
populations inferred from pyrosequencing to the clonal sequences,
we use the measure $\pc_1$ which 
indicates the percent of the inferred population that
matches a clonal sequence within one amino acid difference.
This is used instead of $\pc_{10}$ used in Section~\ref{sec:sim}
to provide a more sensitive performance measure.
In all samples, at least 51.8\% of the inferred populations were within one
amino acid difference of a clonal haplotype.  Based on the present data, we
cannot decide whether the additional inferred haplotypes went undetected by the
Sanger sequencing, or if they are false positives of the reconstruction method.

We found many additional haplotypes in our analysis of the most complex sample,
V11909.  Table~\ref{tab:pop} shows a comparison between the inferred population
for V11909 and the clonal haplotypes.  The populations were analyzed at 15
positions in the protease associated with drug resistance, taken from the HIV
Drug Resistance Database \cite{Rhee2003}.  All but 4 of the 65 clonal
haplotypes (6.1\%) are matched in the inferred population, and the frequencies
in the inferred population are a reasonable match to the frequencies of the
mutation patterns in the clonal haplotypes.  Using the Lander--Waterman model,
we find that the pyrosequencing reads obtained from the HIV samples are enough
to reconstruct with 99\% probability all haplotypes that occur at a frequency
of at least 2.2\% (Eq.~\ref{eq:LW}, Tab.~\ref{tab:results}).  
By comparison, the Sanger sequencing approach yielded 65 clonal sequences,
37 of which were mixtures of two or more clones.

\begin{table}[tb]
        \centering
        \begin{tabular}{lp{.2cm}ccccp{.2cm}cccp{.2cm}ccc}
                \hline
                &&\multicolumn{4}{c}{Pyrosequencing} &&
                \multicolumn{3}{c}{Haplotype reconstruction} &&
                \multicolumn{3}{c}{Comparison to clonal seq.\ }\\
                \cline{3-6} \cline{8-10} \cline{12-14}
                Sample && Reads & Irred  & $\rho_{99}$ & Gaps/kb &  & Err/kb & Min.\ cover &
                Diversity && Clones & Avg.\ dist.\ & $\pc_1$\\
                \hline
                V11909 && 5177 & 641 & 2.2 & 3.3 && 1.10 & 22& 15.8 && 65 & 1.81 & 51.8\\
                V54660 && 7777 & 228 & 1.5 & 2.3 && 1.67 & 4 & 1.0  && 32 & 0.34 & 99.6\\
                V3852  && 4854 & 227 & 2.4 & 3.4 && 1.33 & 7 & 1.4  && 42 & 0.29 & 100\\
                V2173  && 6304 & 354 & 1.8 & 2.3 && 1.31 & 4 & 2.3  && 26 & 0.81 & 86.6\\
                \hline
                ~\\
        \end{tabular}
        \caption{Population reconstruction from four HIV samples.
		The first four columns
        describe the pyrosequencing data: the number of reads,
        the number of irredundant reads (see Theorem~\ref{thm:dilworth}), 
        the expected frequency (in percent) of the least frequent haplotype
        we can expect to cover with 99\% confidence, and 
        the number of gaps/kb in  the aligned data.  The next three columns
        describe the reconstruction algorithm:  the number of non gap
        characters changed in error correction, the size of a
        minimal explaining set of haplotypes, and the diversity, measured as the
        expected number of amino acid differences among the estimated population.
        After error correction, reads were translated into amino acids.
        The final 3 columns describe the validation using (translated) clonal
        sequences: the number of clones sequenced, 
        the average distance between the estimated population and the closest
		Sanger haplotype, and the percent ($\pc_1$) of the estimated population
        that was close (up to 1 amino acid difference) to a clone. }
        \label{tab:results}
\end{table}


\begin{table}[tbp]
        \centering
\begin{tabular}{r@{~~~~~}r@{~~~~~}l}
\hline
\multicolumn{2}{c}{Frequency} \\
\cline{1-2}
Sanger & Pyro & \multicolumn{1}{c}{Mutations} \\\hline
52.3 &  19.3     & M46I, I54V, G73I, I84V, L90M\\
12.3 &  19.0     & M46I, I54V, G73S, I84V, L90M\\
9.2 &   9.4      & M46I\\
6.2 &   5.6      & \\
4.6 &   7.1      & M46I, I54V, G73S, L90M\\
4.6 &   1.9      & M46I, I54V, G73I, L90M\\
3.1 &   5.8      & M46I, G73I, I84V, L90M\\
3.1 &   0.0      & L33F, M46I, I54V, G73S, I84V, L90M\\
1.5 &   1.9      & M46I, L90M\\
1.5 &   0.0      & L33F, M46I, I54V, G73I, I84V, L90M\\
1.5 &   0.0      & M46I, I54V, G73N, I84V, L90M\\
0.0 &   4.9      & M46I, G73I\\
0.0 &   4.7      & M46I, I84V\\
0.0 &   4.0      & M46I, G73S, I84V, L90M\\
0.0 &   3.1      & M46I, I54V, G73S, V82I, I84V, L90M\\
0.0 &   2.9      & M46I, I54V, G73I, I84V\\
0.0 &   2.9      & M46I, I50V, I54V, G73I, I84V, L90M\\
0.0 &   2.0      & I84V\\
0.0 &   1.4      & I54V, G73I, I84V, L90M\\
0.0 &   1.2      & M46I, I50V, I54V, G73S, L90M\\
0.0 &   1.1      & M46I, I54V\\
0.0 &   1.0      & M46I, I50V, I84V, L90M\\
0.0 &   0.5      & M46I, I54V, G73I\\
0.0 &   0.3      & G73S, I84V, L90M\\
\hline
~\\
\end{tabular}
\caption{Comparison of the patterns of resistance mutation
for the 65 Sanger sequences and the 
estimated population for sample V11909.  Mutation patterns were
restricted to 15 positions in the protease
(amino acids
23, 24, 30, 32, 33, 46, 48, 50, 53, 54, 73, 82, 84, 88, and 90)
associated with PI resistance.}
\label{tab:pop}
\end{table}

\section{Discussion}
\label{sec:discuss}

Pyrosequencing constitutes a promising approach to estimating the genetic
diversity of a community. However, sequencing errors and short read lengths
impose considerable challenges on inferring the population structure from a set
of pyrosequencing reads.  We have approached this task by identifying and
solving consecutively three computational problems: error correction, assembly
of candidate haplotypes, and estimation of haplotype frequencies.  Our methods
focus on the situation where a reference genome is available for the alignment
of reads. This is the case, for example, for many important pathogens, such as
bacterial and viral populations.

The procedure consists of three steps.  First, error correction is performed
locally.  We take windows of fixed width over the aligned reads  
and cluster reads within the windows in order to resolve the local haplotype
structure.  This approach is a combination of methods
\cite{Kececioglu2001,Tammi2002} specially tailored to pyrosequencing reads.
Next, haplotypes are reconstructed using a new application of a classic
combinatorial algorithm.  This step is the main theoretical advance in this
paper.  Finally, haplotype frequencies are inferred as the ML estimates of a
statistical model that mimics the pyrosequencing process. 
We have developed an EM algorithm for solving this ML problem.

Haplotype reconstruction is based on two assumptions: consistency and
parsimony.  We require that each haplotype be constructible from a sequence of
overlapping reads and that the set of explaining haplotypes be as small as
possible.  The Lander--Waterman model of sequencing implies lower bounds on the
number of reads necessary to meet the first requirement.  The fundamental
object for haplotype reconstruction is the read graph. A minimal set of
explaining haplotypes corresponds to a minimal path cover in the read graph,
and this path cover can be found efficiently using combinatorial optimization.
Moreover, the cardinality of the minimal path cover is an important invariant
of the haplotype reconstruction problem related to the genetic diversity of the
population.  

We believe that these methods are also applicable to many metagenomic projects.
In such projects, estimation of the diversity of a population is a fundamental
question.  The size of a minimal cover of a fragment assembly graph provides an
intuitive and computable measure of this diversity.

We have validated our methods by extensive simulations of the pyrosequencing
process, as well as by comparing haplotypes inferred from pyrosequencing data
to sequences obtained from direct clonal sequencing of the same samples.  Our
results show that pyrosequencing is an effective technology for quantitatively
assessing the diversity of a population of RNA viruses, such as HIV.

Resistance to most antiretroviral drugs is caused by specific mutational
patterns comprising several mutations, rather than one single mutation. Thus,
an important question that can be addressed efficiently by pyrosequencing is
which of the resistance mutations actually occur on the same haplotype in the
population.  Since our methods avoid costly clonal sequencing of the HIV
populations for determining the co-occurrence of HIV resistance mutations
\cite{Gonzales2003c},   pyrosequencing may become an attractive alternative to
the traditional clonal Sanger sequencing.

The sample size of approximately  10,000 reads we have considered provides us
with the opportunity of detecting variants present in only 1\% of the
population.  Pyrosequencing can produce 200,000 reads and thus twenty
populations could be sequenced to a good resolution using a process less labor
intensive than a limiting dilution clonal sequencing to a similar resolution of
a single population.

The simulations suggest that the method works best with populations that are
suitably diverse.  Intuitively, the information linking two reads together on
the same haplotype decays rapidly in sections of the genome where there are few
identifying features of that haplotype (as in a region of low diversity).  In
particular, repeats of sufficient length in the reference genome can completely
destroy linkage information.  However, at some point the benefits of increased
diversity will be partially reduced by the increased difficulty of the
alignment problem.  With more diverse populations or true indels,
alignment to a single reference genome will become less accurate.

The HIV {\em pol} gene analyzed
here is on the low end of the diversity spectrum.  The {\em env} gene with its
higher variability may be a better target for some applications.  We expect the
proposed methods to improve early detection of emerging drug resistant variants
\cite{Doukhan2001,Metzner2005}, and to support the genetic and epidemiological
study of acute infections, in particular the detection of dual infections
\cite{Gottlieb2004}.

Since our computational procedure produces an estimate of the entire virus
population, it allows the study of fundamental questions about the evolution of
viral populations in general \cite{Rambaut2004}. For example, mathematical
models of virus evolution can be tested directly within the accuracy of
estimated viral haplotype frequencies \cite{Rouzine2001}.  Predicting viral
evolution is considered an important step in HIV vaccine development
\cite{Gaschen2002}.

In addition to the promising biological applications, there are many
interesting theoretical questions about reconstructing populations from
pyrosequencing data.
The errors in pyrosequencing reads tend to be highly correlated, as they occur
predominately in homopolymeric regions.  While this can make correction more
difficult (a fact which can be counteracted by the use of quality scores), we
believe that it can make haplotype reconstruction more accurate than if the
errors were uniform.  If errors are isolated to a few sites in the genome,
fewer additional explaining haplotypes are needed than if the errors were
distributed throughout.  The exact relationship between the error process of
pyrosequencing, error correction, and haplotype reconstruction is worthy of
further study.

As pyrosequencing datasets can contain 200,000 reads, it is worthwhile to
investigate how our methods scale to such large datasets.  
Haplotype reconstruction is the only step which is not immediately practical on
such a large number of reads, since it is at worst cubic in the number of
irredundant reads.  However, problems of this size are  approachable
with our methods as follows.

The theoretical resolution of the algorithms depends on two factors: first, the
ability to differentiate between errors and rare variants; and second, whether
there are enough reads so that we can assemble all haplotypes.  We have seen
that the number of reads necessary for assembly scales with the inverse of the
desired resolution (Eq.~\ref{eq:LW}):  if $N$ reads cover all haplotypes of frequency at least
$\rho$, then $kN$ reads are needed to cover all haplotypes of frequency at
least $\rho/k$.  However, the resolution of error correction is at most the
overall error rate as the number of reads grows;  
see Table~\ref{tab:ecres}.

\begin{table}
	\centering
	\begin{tabular}{lrrrrrr}
		\hline
		Number of reads  & 1000 & 5000 & 10,000 & 50,000 & 100,000 & 200,000\\
		\hline
		Error resolution(\%) & 3 & 1.2 & 0.9 & \underline{0.5} & \underline{0.42} & \underline{0.365}\\
		Reconstruction resolution (\%)& \underline{11.5} & \underline{2.3} & \underline{1.2} & 0.23 & 0.12 & 0.058\\
		\hline
		~
	\end{tabular} 
	\caption{Resolution of the error correction (binomial test only) and
	haplotype reconstruction as a function  of number of reads.  The resolution
	of the error correction is defined as the smallest frequency of a mutation
	that will be visible over the background error rates; it is calculated with
	error rate $\epsilon = 0.0025$ and significance level $\alpha = 0.001$.
	The resolution of the haplotype reconstruction (derived from the
	Lander--Waterman model, (Eq.~\ref{eq:LW})) is the  smallest haplotype
	frequency expected to be entirely covered by reads.  For small read sizes,
	the
	haplotype reconstruction is the limiting factor (underlined in the table)
	but for over approximately $35,000$ reads  the error correction is the limiting
	factor.}
	\label{tab:ecres}
\end{table}

The limited resolution of error correction combined with the elimination of
redundant reads makes haplotype reconstruction feasible for large datasets.
For example, error correction on 200,000 reads with $\epsilon = 0.0025$ and
$\alpha = 0.001$ will erase all variants with frequency below $0.365\%$ (
Table~\ref{tab:ecres} and Section~\ref{sec:ecmethod}).  In order to have enough
information to reconstruct these variants under the Lander--Waterman model, we
would expect to need  only about 30,000 reads.  Furthermore, in regions of
low diversity, many of the reads will be redundant and are thus discarded
before building the graph.  For example, with 30,000 error-free reads simulated
from $275 \approx 1/0.00365$ haplotypes at 5\% diversity, typically about
13,000 reads are irredundant.  This number of irredundant reads is near the
limits of our current implementation.

Current and future improvements to pyrosequencing technology will lead to
longer reads (250 bp), more reads, and lower errors.  However, in
order for huge numbers of reads to be of great help in the ultra-deep
sequencing of a population, the error rates must also decrease.  The
performance of our methods as read length varies is an important question,
given the availability of sequencing technologies with different read lengths
(e.g., Solexa sequencing with 30--50 bp reads) and the desire to assemble
haplotypes of greater size (e.g., the 10 kb entire HIV genome).  

Notice that haplotype reconstruction seems to be quite good locally (cf.\
Fig.~\ref{fig:RG} and \ref{fig:multCD}) in that many reconstructed haplotypes
contain large contiguous regions where they agree with a real haplotype.  
However, the measures of performance considered in this paper all deal with the
entire haplotype and ignore partial results of this type.
This would seem to imply that longer reads will improve the reconstruction
performance on a fixed length genome; new performance measures will have to be
developed to analyze these problems.



%




\section{Methods}

\subsection{Statistical tests for error correction}
\label{sec:ecmethod}
We use two statistical tests for locally detecting distinct haplotypes.  The
first test analyzes each column of the multiple alignment window.  Write $d$
for the number of reads that overlap this window.  We ask if the observed
number of mutations (deviations from the consensus base) exceeds our
expectation under the null hypothesis of one haplotype and a uniform sequencing
error $\epsilon$.  The probability of observing $x$ or more mutations is given
by the binomial distribution 
\begin{equation}
	\label{eqn:binom}
   \Pr(X \ge x) = \sum_{k=x}^d{d \choose k} \epsilon^k (1-\epsilon)^{d-k}.
\end{equation}
There are two parameters to set here: the error rate $\epsilon$ and the
$p$-value $\alpha$ that is required for significance.

Next, we test pairs of mutations in two different alignment columns $u$ and $v$
using Fisher's exact test.  The test statistic is the number $C$ of
co-occurrences, which under the null hypothesis of one haplotype follows the
hypergeometric distribution 
\begin{equation}
	\label{eqn:hyperg}
   \Pr(C = c) = \frac{{n_v \choose c} 
     {d - n_v \choose n_u - c}}{{d \choose n_u}}, 
 \end{equation}
where $n_u$ and $n_v$ are the number of times the specific mutations have been
observed in columns $u$ and $v$, respectively \cite{Tammi2002}.  
Considering pairs provides more power if co-occurrences are observed on reads,
but cannot detect single mutation differences.  We set the $p$-value for this
test to be the same as for the binomial test.

The procedure tests all columns in the window using (\ref{eqn:binom}) and then
all  pairs of columns using (\ref{eqn:hyperg}).  This can lead to over-counting
of the number of haplotypes as follows. Suppose that in columns 1 and 2 the
consensus base is $\A$, but that there is a mutation $\C$ in some of the reads
in each column.  If both mutations are significant by themselves, this is
evidence of three haplotypes in the window.  If they are also significant
together, this would be evidence of four haplotypes. However, there could be
only two true haplotypes  at these two positions: \A\A\  and \C\C.  To
correct for this, we subtract two from the count whenever two significant
mutations are significant together and always occur on exactly the same set of
reads.  


We do not explicitly address the multiple comparisons problem associated with
this testing procedure here and regard the significance levels of the tests as
parameters of Algorithm~\ref{alg:ec}.  We account for the quality scores
associated with each base by using (rounded) weighted counts in the test
statistics.  Gaps are treated as unknown bases and represented by a special
character with quality score zero.  We found that an error rate of $0.0025$, a
$p$-value of $0.001$, and a window size of $24$ provided the best error
correction.  These parameters can be tuned as follows.  First, the window
size should be chosen to best help the clustering.  A large window provides
more power since there are more identifying mutations, but also can be more
difficult to cluster since many reads will only partially overlap the window.

Next, the $p$-value for the tests and the error rate should be adjusted to
prevent false positives and negatives.  The number of mutations
required in a column before the mutation is considered significant
can be calculated from (\ref{eqn:binom}).
For example, with 10,000 reads of length 100 in a genome of length 1000, there
will be approximately $d = 1000$ reads overlapping a small window.  Setting
$\epsilon = 0.0025$ and $\alpha = 0.001$, a mutation would have to occur nine
times in a column to be significant according to (\ref{eqn:binom}).  Thus the
error correction would (roughly speaking) discard any mutations occurring in
less than $9/1000 \approx 1\%$ of the population.  Notice that this is quite
similar to the estimate under the Lander--Waterman model (Eq.~\ref{eq:LW}),
where 11,508 reads are needed to cover all haplotypes at 1\% frequency.

Notice that the power of the error correction grows very slowly, see
Table~\ref{tab:ecres}.  On a dataset with 200,000 reads, then error correction
would eliminate any variants present in less than $0.365\%$ of the population.
However, only $30,000$ reads are needed to achieve this resolution with
haplotype reconstruction.

\subsection{Proof of Theorem~\ref{thm:dilworth}}
\label{sec:proof}

Suppose the read graph $\RG{\sR}$ has $V$ vertices and $E$ edges.  Since
$\RG{\sR}$ is acyclic, it defines a partial order on the set of irredundant
reads, $\sR_{\rm irred}$.  Part (1) is then a direct application of
Dilworth's theorem \cite{Dilworth1950} to this partially ordered set.

The associated bipartite graph has vertex set $\{A, B \}$, where both $A$ and $B$ are
equal to $\sR_{\rm irred}$.  There is an edge between $r \in A$ and $s
\in B$ if there is a path from $r$ to $s$ in $\RG{\sR}$.  Then a maximal
matching in the bipartite graph is equivalent to a minimal chain decomposition
of $\RG{\sR}$ (see \cite{Ford1962}).  

For the time complexity, 
notice that building the read graph $\RG{\sR}$ is of complexity $O(V^2)$.
Building the associated bipartite graph is
equivalent to finding the transitive closure of the read graph and thus is
$O(V E)$.  The efficient matching algorithm for the solution of the
matching problem is due to Hopcroft and Karp \cite{Hopcroft1973}.  For a
general bipartite graph with $V'$ vertices and $E'$ edges, 
the Hopcroft-Karp algorithm is of time
complexity  $O(E'\sqrt{V'})$.  Since in our construction, $V' = 2V$ and
$E' = O(V^2)$, the matching algorithm takes time $O(V^{5/2})$.
Depending on the structure of the graph, either the transitive closure
or matching problems can dominate, but both are of complexity $O(V^3)$.

\subsection{EM algorithm for haplotype frequency estimation} 
\label{sec:emmethod}
We use an EM algorithm \cite{Dempster1977} to estimate the maximum likelihood
haplotype frequencies.  We iteratively
estimate the missing data $u_{rh}$, i.e., the 
number of times read $r$ originated from haplotype $h$,
and solve the easier optimization problem of 
maximizing the log-likelihood of the hidden model
\[
   \ell_{\rm hid}(p_1, \dots, p_{|\sH|}) = 
     \sum_{r \in \sR} \sum_{h \in \sH} u_{rh}
     \log \left( p_h \Pr(R = r \mid H = h) \right).
\]
In the E step, the expected values of the missing data are 
computed as
\[
   u_{rh} = u_r \frac{p_h \Pr(R = r \mid H = h)}{\Pr(R = r)}.
\]
In the M step, maximization of $\ell_{\rm hid}$ yields
\[
   \hat{p}_h = \frac{\sum_{r \in \sR} u_{rh}}{\sum_{r \in \sR} u_r}.
\]

\subsection{Simulations} \label{sec:methodsim}

Our starting point is the first 1kb of the wild type sequence of the HIV {\em pol}
gene, encoding the 99 amino acids of the protease and the beginning of the
reverse transcriptase. Random mutations are introduced into this strain in
order to generate genetic diversity. 
We generated various populations in this way with diversities between 20 and 80
base pairs (2--8\%). 
All haplotypes were set to have the same frequency in the population.  

We report the expected value of the Hamming distance 
between two haplotypes drawn from a population as our basic measure of the
diversity of a population.  This statistic, which we call simply ``diversity''
can be thought of as a version of the Simpson measure \cite{Simpson1949} that
takes into account the genetic structure.  

We use ReadSim \cite{Schmid2006} 
(available from \url{http://www-ab.informatik.uni-tuebingen.de/software/readsim/welcome.html})
to simulate the error process of
pyrosequencing.  We generate reads by running ReadSim with the options
\texttt{-meanlog  0.15 -sigmalog 0.08 -filter} (aside from these options, we use the default parameters). 
This process results in about 7 insertions and 3 deletions per kb.  
Since pyrosequencing produces light coverage on the
tails of the input genomes, we simulate by padding the region of concern with 100
nucleotides on each end and discard reads from the tails.
The error correction (Algorithm~\ref{alg:ec}) is run with window size of 24,
$p$-value of 0.001, and error rate $\epsilon = 0.0025$.  
We recorded  the frequencies of error at each position in the genome during
simulations of populations of size 10 at 5\% diversity.  Sampling from 
these frequencies allowed us to create reads with precise error rate
for Fig.~\ref{fig:close}.

For the simulations of haplotype reconstruction, we generate
pyrosequencing reads using the model described in Section~\ref{sec:em}
and illustrated in Fig.~\ref{fig:sampling} 
complemented by uniform sequencing errors
at rate $0$, $0.1$, and $0.2$ per kb.
We build the read graph and apply Algorithm~\ref{alg:min}
repeatedly until 200 candidate haplotypes are found.  
The EM algorithm was run with 10 random starting points.  
To speed up the EM algorithm, we round all
frequencies $p_h < 10^{-6}$ to zero.

We also test a simple alternative to the EM algorithm as follows.  For each
haplotype $h \in \sH$, let $c_h$ count how many reads haplotype $h$ is consistent with
and set $p_h = c_h / \sum_{l \in \sH} c_{l}$.  This estimate will
be correct under the given model if each read is consistent with exactly
one haplotype.

For evaluating the performance of the various steps of our reconstruction
method, we use several basic measures of performance.  To measure the distance
between two sets of haplotypes (one original and one inferred), we calculate
how many of the original haplotypes are found among the inferred haplotypes as
well as the average of the distances between each inferred haplotype and its
closest original haplotype (Fig.~\ref{fig:recons}).
Distance is measured as Hamming distance on the amino acid level.
To compare two populations with different frequencies but the same
haplotypes, we use the Kullback--Leibler (KL) divergence $D_{\rm KL} (p
\parallel q) = \sum_{h\in \sH} p_h \log (p_h / q_h)$, where $p$ and $q$ are the
two discrete (haplotype) distributions with the same support $\sH$ (Fig.~\ref{fig:EM}).
To measure the performance of the entire process, we measure how much of
the inferred population is close to the original population.
Specifically, we calculate the percent of the inferred population that is
within a specified distance from one of the original haplotypes
(cf.\ Fig.~\ref{fig:close}).  
We refer to this statistic as $\pc_n$, where $n$ is the number of amino acid differences we allow.

\subsection{HIV sequence data}

Virus populations derived from four treatment-experienced patients
between 2000 and 2005 were sequenced using both
pyrosequencing and limiting dilution Sanger sequencing.
The plasma HIV-1 RNA levels in the four
plasma samples were each greater than 100,000 copies/ml as determined 
using the VERSANT HIV-1 RNA assay \cite{Collins1997}.
Each sequence encompassed all 99 HIV-1 protease codons 
and the first 241 reverse transcriptase codons.
The same genomic region of the same four samples was analyzed 
using limiting dilution and direct Sanger sequencing of the clones.
Sample preparation and pyrosequencing and Sanger sequencing techniques
are explained in detail in \cite{Wang2007}.  

Briefly, ultra-deep pyrosequencing was performed
on four RT-PCR products from RNA extracted from cryopreserved plasma samples. 
The median number of cDNA copies prior to sequencing was 100 with an 
interquartile range of 75 to 180.  The resulting datasets consisted of 
between 4854 and 7777 reads of average length 105 bp.
Reads were error corrected (Algorithm~\ref{alg:ec}) and translated 
to amino acids. For haplotype reconstruction, Algorithm~\ref{alg:min}
was run repeatedly until all or at most 10,000 candidate haplotypes
were found.
The samples were translated into amino acids after the error correction step; 
thus, the haplotype reconstruction and frequency estimation algorithms are 
done on the amino acid level.


For the sample with the greatest diversity (V11909), the unamplified cDNA product was serially
diluted prior to PCR amplification. 
Bidirectional sequencing was
performed directly on 37 amplicons derived from the 1/30 cDNA dilutions and 31 amplicons
derived from the 1/100 cDNA dilutions.   Three sequences were discarded because of incomplete coverage.
For the other three samples, we used the Sanger method to sequence a total of
32, 42, and 26 plasmid subclones per sample.

Some of the sequences obtained from limiting dilutions contained mixtures of several
clones.  In this case, in order to measure the Hamming distance between an inferred
haplotype and a clonal haplotype with ambiguous bases, we used the minimum distance
over all possible translations of the ambiguous haplotype.


\section*{Acknowledgments} 
N.~Eriksson and L.~Pachter were partially supported
by the NSF (grants DMS-0603448 and CCF-0347992, respectively).  
N.~Beerenwinkel was funded by a grant from the Bill \& Melinda
Gates Foundation through the Grand Challenges in Global Health
Initiative.

\bibliographystyle{plos}

\newpage

\begin{appendix}

\setcounter{section}{19}  
\setcounter{figure}{0}
\renewcommand{\thefigure}{\thesection\arabic{figure}}

\section*{Supporting Information}

\begin{figure}[h!]
	\centering
	\includegraphics[width=\textwidth,height=.35\textheight]{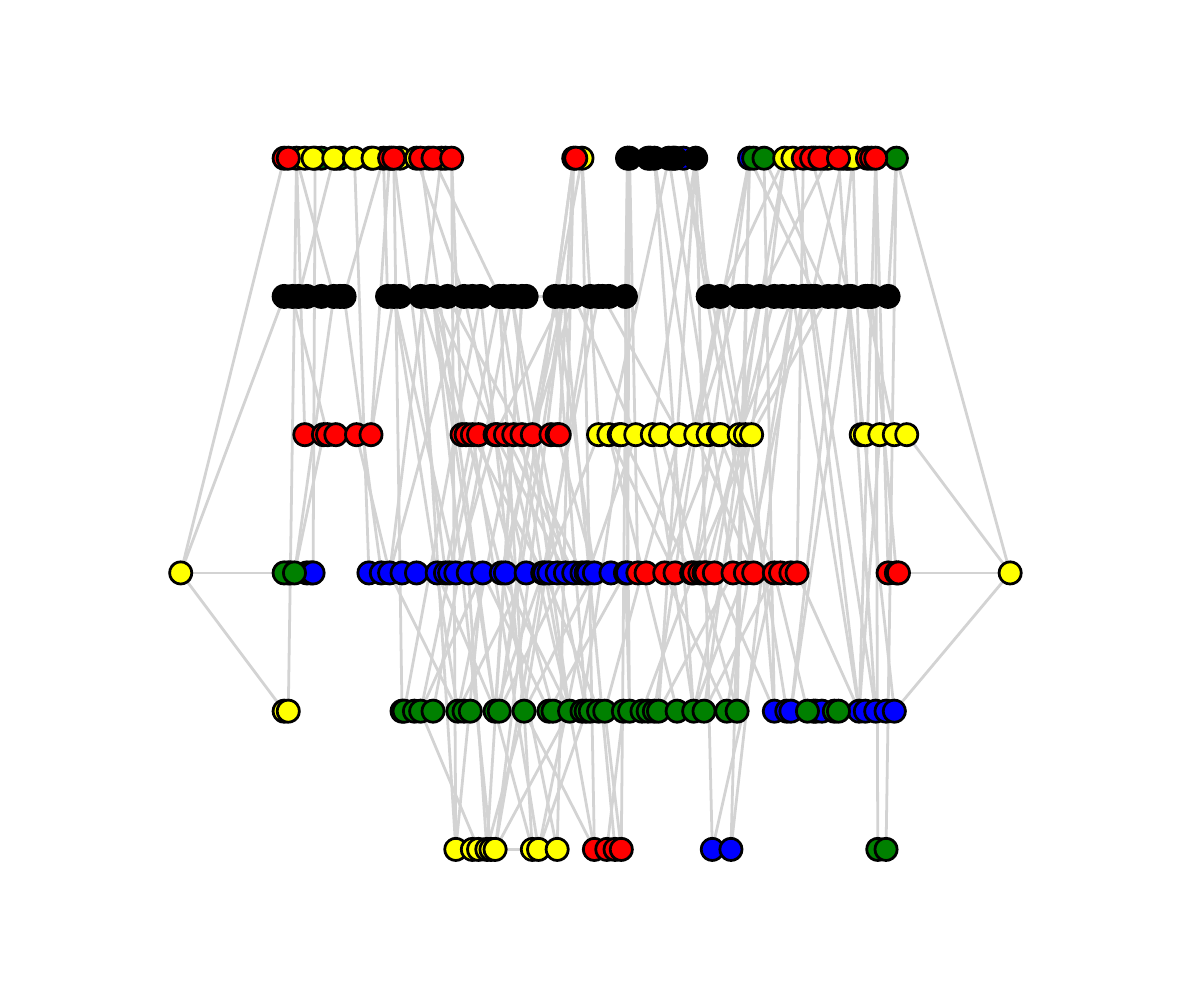}
	\includegraphics[width=\textwidth,height=.35\textheight]{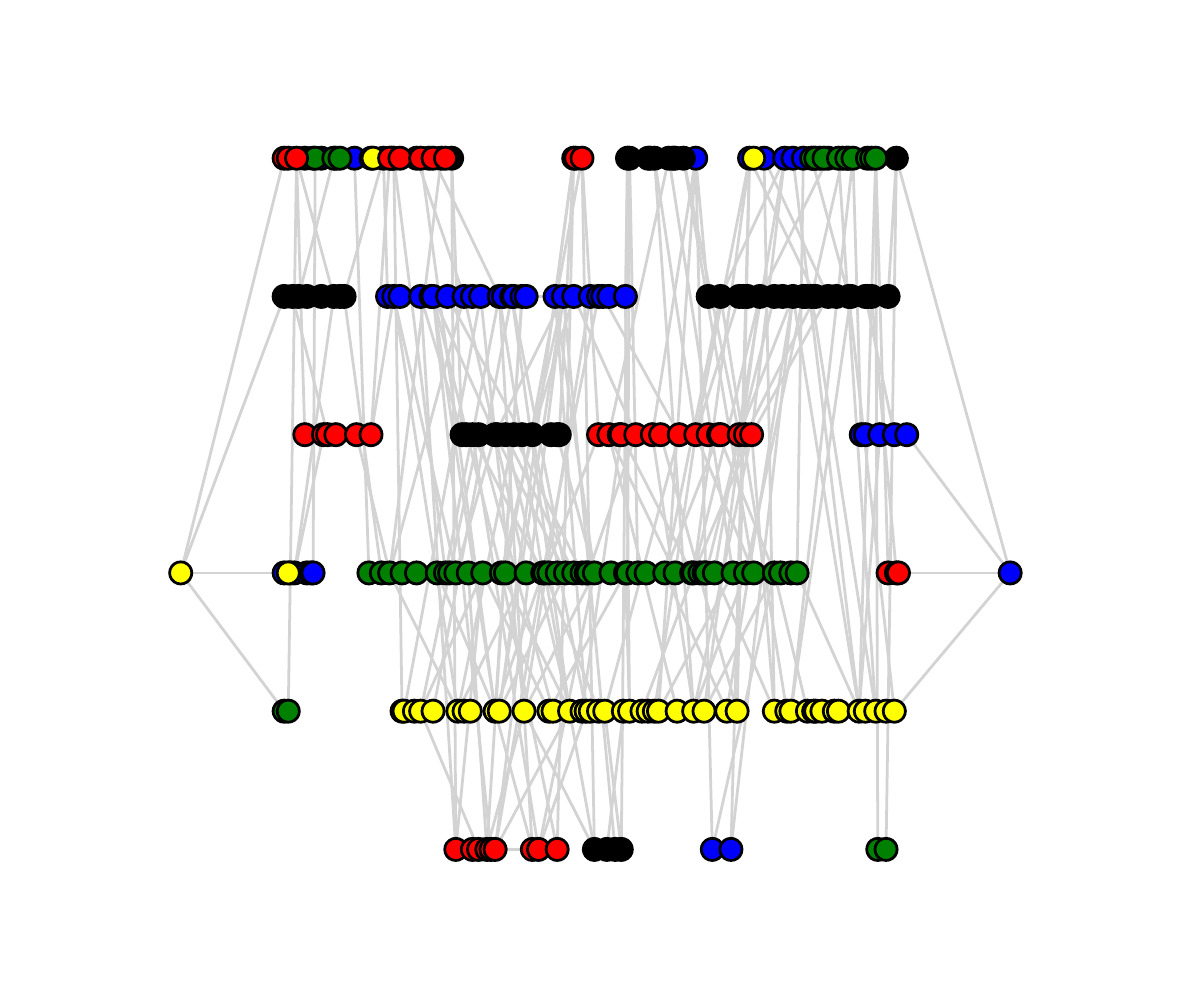}
	\caption{Two different chain decompositions of the read graph for 1000
	reads from a population of 5 haplotypes at 3\% diversity. The bottom five
	lines correspond to reads matching a haplotype uniquely; the top to reads
	matching several haplotypes.  One decomposition gets one
	haplotype entirely correct (top, black); the other gets two different
	haplotypes essentially correct (bottom, green and yellow).  In this way,
	taking multiple chain decompositions allows us to reconstruct all
	haplotypes.}
	\label{fig:multCD}
\end{figure}

\begin{figure}[h!]
	\centering
	\begin{tabular}{c@{\hspace{1cm}}c}
		\includegraphics[width=.45\textwidth]{figures/10haplo-dist} & 
		\includegraphics[width=.45\textwidth]{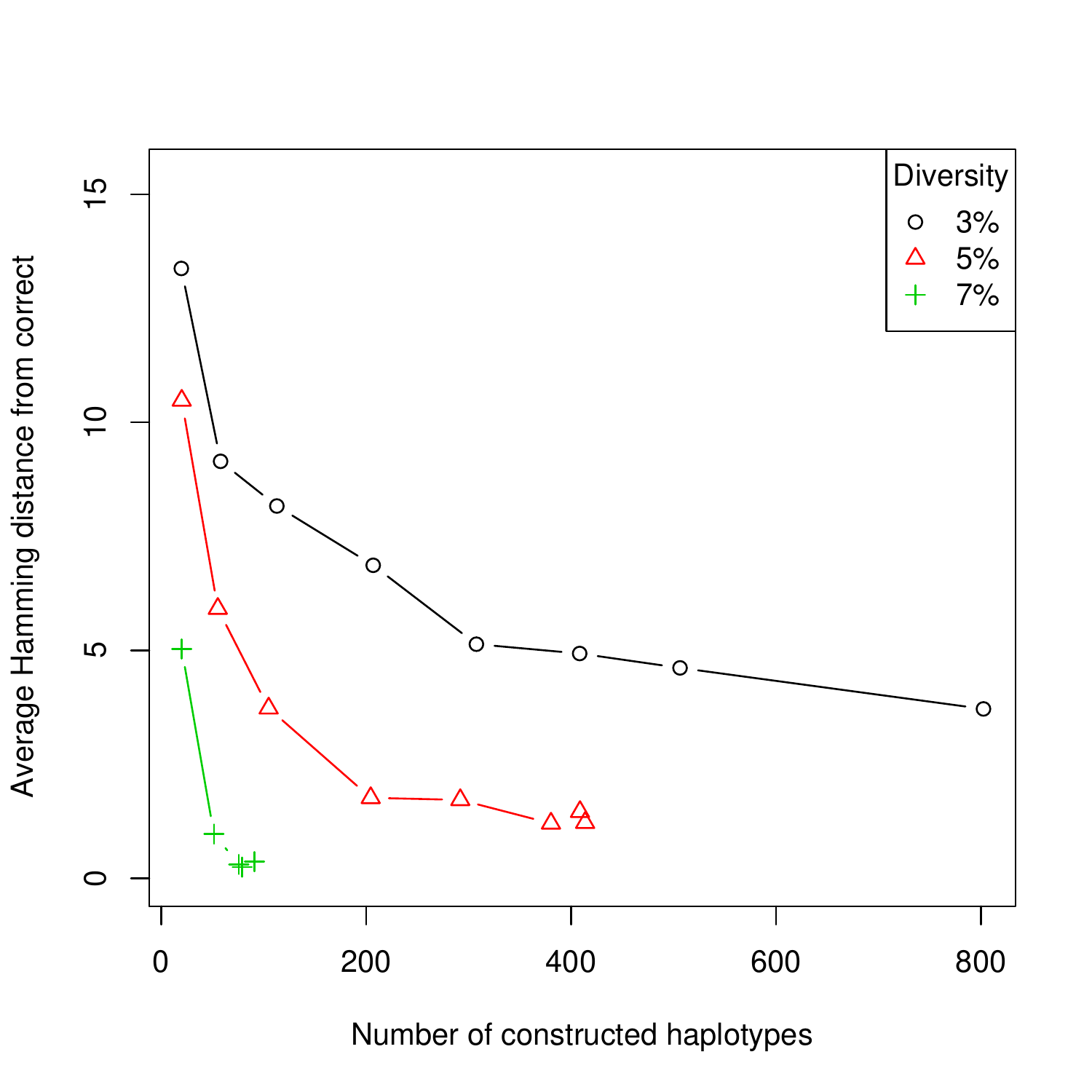}\\
		(a) & (b) \\
	\end{tabular}
	\includegraphics[width=.45\textwidth]{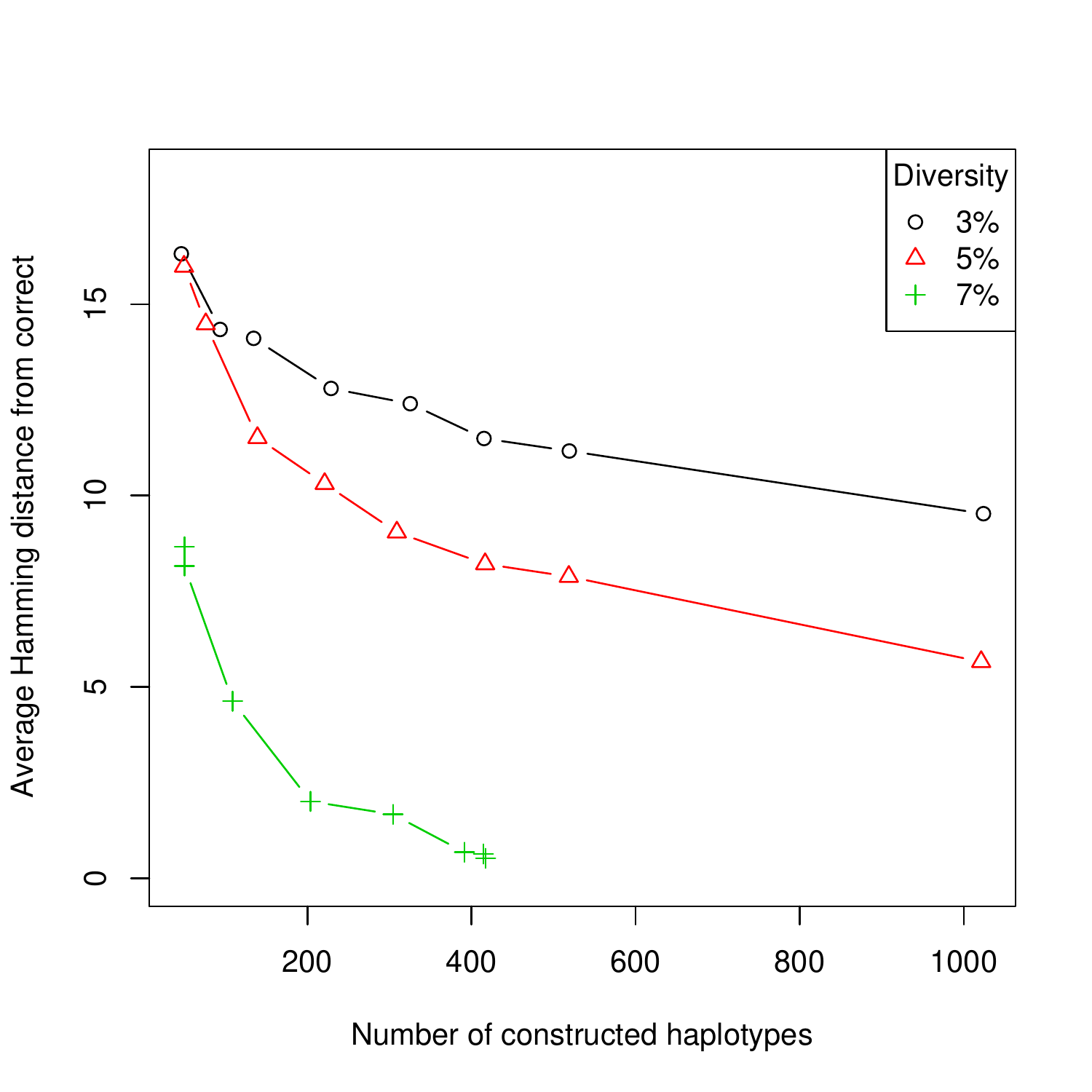} \\
	(c)
	\caption{Haplotype reconstruction.  Up to 1000 candidate haplotypes were
	generated using Algorithm~\ref{alg:min} from 10,000 error free reads drawn
	from populations of size 10, 20, and 50 (subfigures (a), (b), and (c))
	at varying diversity levels.  Displayed is a measure of the efficiency of
	haplotype reconstruction: the average Hamming distance (in amino acids)
	between an original haplotype and its closest match among the reconstructed
	haplotypes. }
	\label{fig:20haplo}
\end{figure} 

\begin{figure}[h!]
	\centering
	\includegraphics[width=\textwidth]{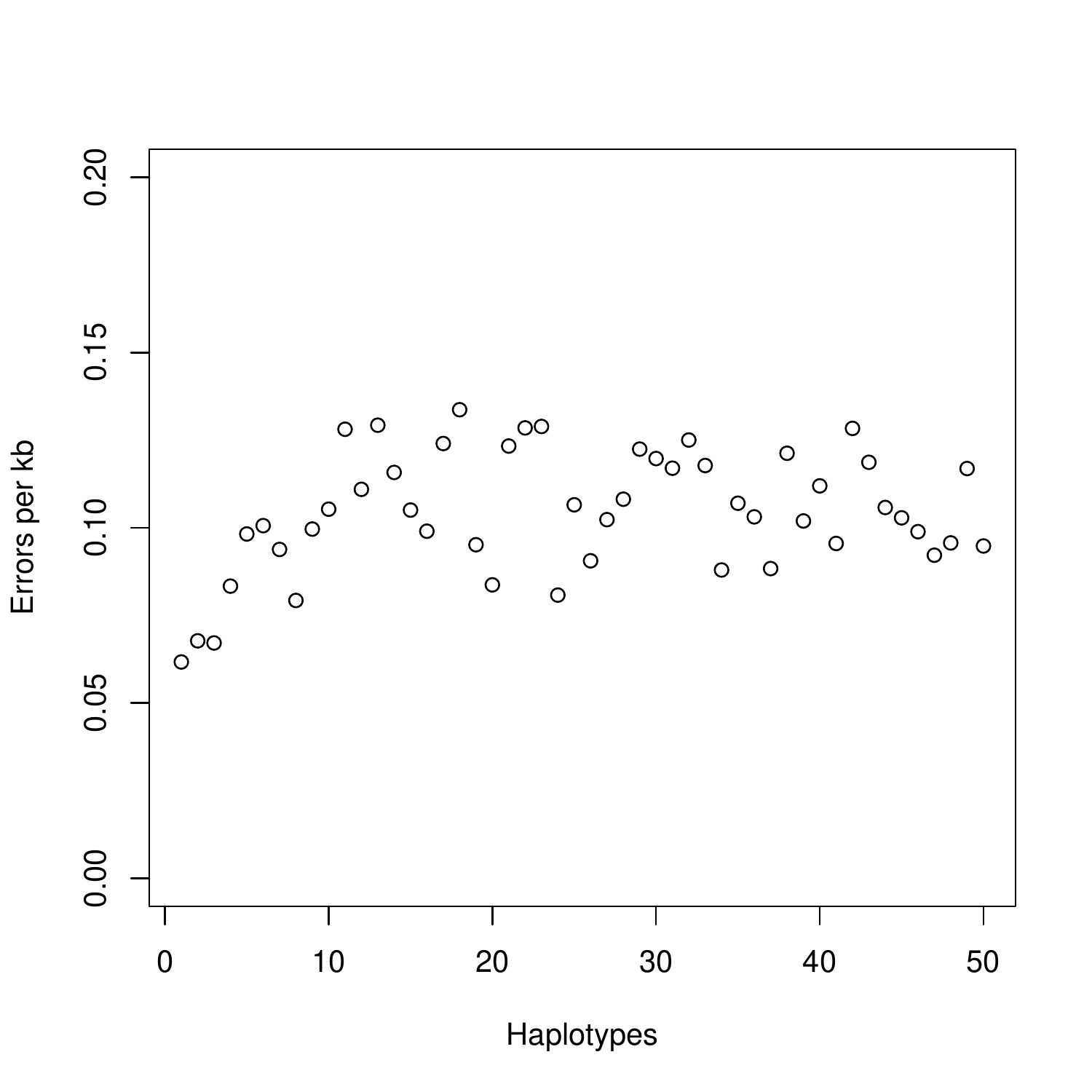} 
	\caption{Resulting error after error correction on populations with 4\%
	diversity.  Populations with up to 50 haplotypes of equal frequency were
	created.  \texttt{ReadSim} was used to simulate pyrosequencing with an
	error rate of 3--6 errors per kb (after alignment).  Error correction
	sucessfully reduced the error rate by a factor of approximately 30.}
	\label{fig:ec}
\end{figure} 

\begin{figure}[h!] 
	\centering
	\includegraphics[width=\textwidth]{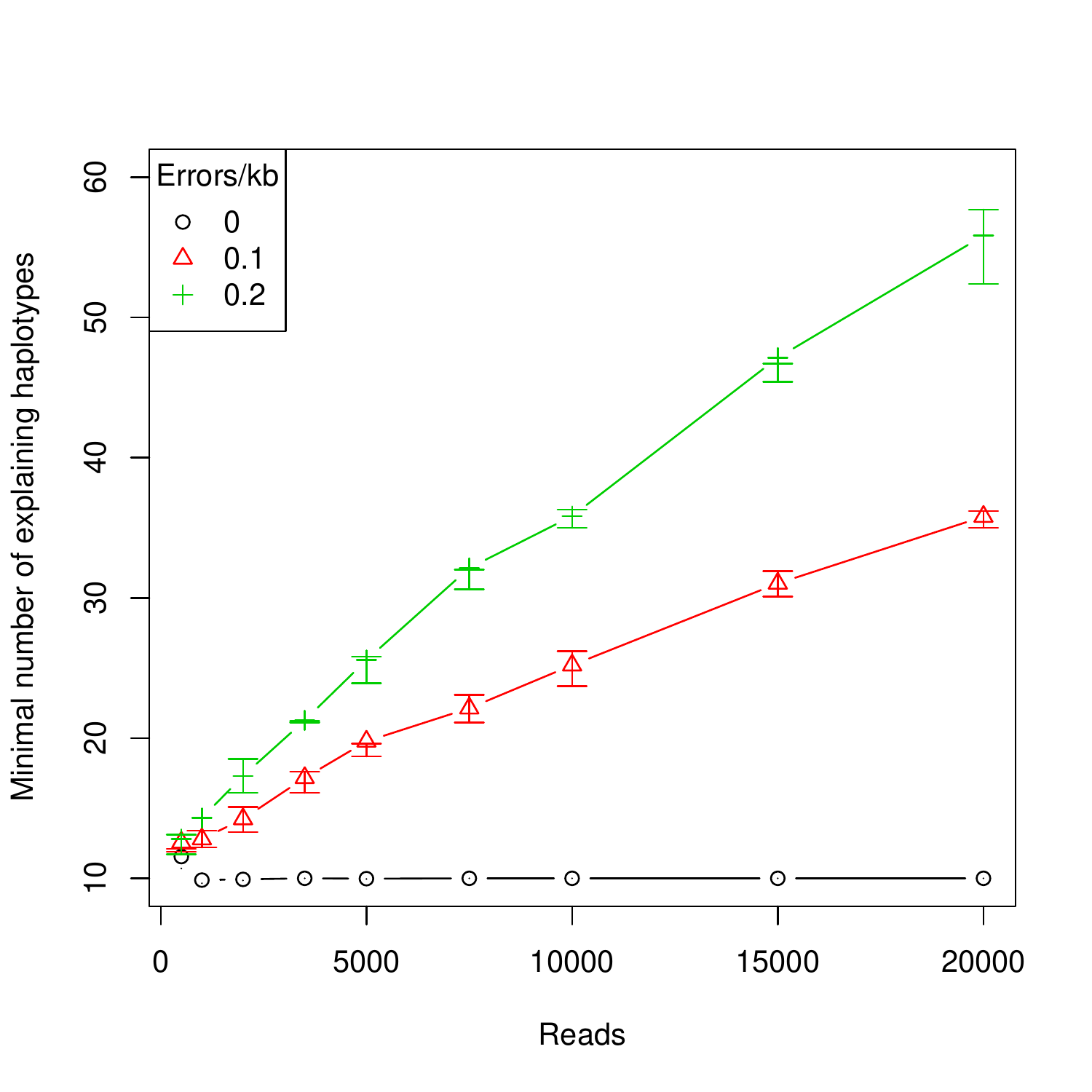}
	\caption{Lower bound on population size for simulations with varying error
	rate and number of reads. Populations ranged from 3-7\% diversity.  The
	lower bound is computed as the minimal size of a cover of the read graph.
	Error bars give interquartile range over 100 trials at different diversity
	levels.  This estimated lower bound is quite accurate for error free reads;
	it seems to increase linearly with the number of reads if errors are
	introduced.}
	\label{fig:antichains}
\end{figure}

\end{appendix}
\end{document}